\documentclass{SciPost}

\hypersetup{
    colorlinks,
    linkcolor={red!50!black},
    citecolor={blue!50!black},
    urlcolor={blue!80!black}
}

\usepackage[bitstream-charter]{mathdesign}
\DeclareSymbolFont{usualmathcal}{OMS}{cmsy}{m}{n}
\DeclareSymbolFontAlphabet{\mathcal}{usualmathcal}

\fancypagestyle{SPstyle}{
\fancyhf{}
\lhead{\colorbox{scipostblue}{\bf \color{white} ~SciPost Physics }}
\rhead{{\bf \color{scipostdeepblue} ~Submission }}

\fancyfoot[C]{\textbf{\thepage}}
}

\usepackage{bm}% bold math
\usepackage{graphicx}
\usepackage{comment}
\usepackage{xcolor}

\begin{document}

\pagestyle{SPstyle}

\begin{center}{\Large \textbf{\color{scipostdeepblue}{
Spin-only dynamics of the multi-species nonreciprocal Dicke model\\
}}}\end{center}

\begin{center}\textbf{
% Use (full) first name (+ middle name initials) + surname format.
% Separate subsequent authors by a comma, omit comma and use "and" for the last author.
% Mark the corresponding author(s) with a superscript symbol in this order
% \star, \dagger, \ddagger, \circ, \S, \P, \parallel, ...
Joseph Jachinowski\textsuperscript{1$\star$} and
Peter B. Littlewood\textsuperscript{1,2$\dagger$}
}\end{center}

\begin{center}
{\bf 1} James Franck Institute and the Department of Physics, University of Chicago, Chicago, IL 60637, USA
\\
{\bf 2} School of Physics and Astronomy, University of St Andrews, St Andrews KY16 9AJ, UK
%%%%%%%%%% END TODO: AFFILIATIONS
% Provide email address of corresponding author(s)
\\[\baselineskip]
$\star$ \href{mailto:email1}{\small jachinowski@uchicago.edu}\,,\quad
$\dagger$ \href{mailto:email2}{\small littlewood@uchicago.edu}
\end{center}

\section*{\color{scipostdeepblue}{Abstract}}
\textbf{\boldmath{%
The Hepp-Lieb-Dicke model is ubiquitous in cavity quantum electrodynamics, describing spin-cavity coupling which does not conserve excitation number. Coupling the closed spin-cavity system to an environment realizes the open Dicke model, and by tuning the structure of the environment or the system-environment coupling, interesting spin-only models can be engineered. In this work, we focus on a variation of the multi-species open Dicke model which realizes mediated nonreciprocal interactions between the spin species and, consequently, a dynamical limit-cycle phase. In particular, we improve upon adiabatic elimination and, instead, employ a Redfield master equation in order to describe the effective dynamics of the spin-only system. We assess this approach at the mean-field level, comparing it both to adiabatic elimination and the full spin-cavity model, and find that the predictions are sensitive to the presence of single-particle incoherent decay. Additionally, we clarify the symmetries of the model and explore the dynamical limit-cycle phase in the case of explicit $\mathcal{PT}$-symmetry breaking, finding a region of phase coexistence terminating at an codimension-two exceptional point. Lastly, we go beyond mean-field theory by exact numerical diagonalization of the master equation, appealing to permutation symmetry in order to increase the size of accessible systems. We find signatures of phase transitions even for small system sizes.
}}

\vspace{\baselineskip}

%%%%%%%%%% BLOCK: Copyright information
% This block will be filled during the proof stage, and finilized just before publication.
% It exists here only as a placeholder, and should not be modified by authors.
\noindent\textcolor{white!90!black}{%
\fbox{\parbox{0.975\linewidth}{%
\textcolor{white!40!black}{\begin{tabular}{lr}%
  \begin{minipage}{0.6\textwidth}%
    {\small Copyright attribution to authors. \newline
    This work is a submission to SciPost Physics. \newline
    License information to appear upon publication. \newline
    Publication information to appear upon publication.}
  \end{minipage} & \begin{minipage}{0.4\textwidth}
    {\small Received Date \newline Accepted Date \newline Published Date}%
  \end{minipage}
\end{tabular}}
}}
}
%%%%%%%%%% BLOCK: Copyright information

% For convenience during refereeing we turn on line numbers:
% \linenumbers

\vspace{10pt}
\noindent\rule{\textwidth}{1pt}
\tableofcontents
\noindent\rule{\textwidth}{1pt}
\vspace{10pt}

% Write your article contents here, starting from first \section.
% An example structure is given below.

\section{Introduction}

\begin{figure}
    \centering
    \includegraphics[trim = {1cm 2cm 0.5cm 1.6cm}, clip, width=0.6\linewidth]{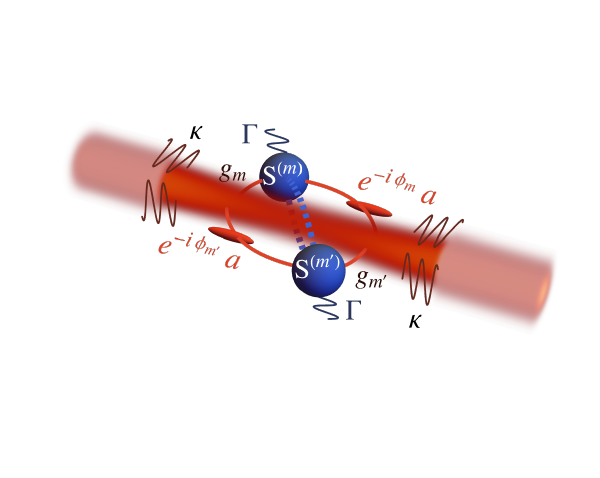}
    \caption{Schematic depiction of the system studied in this paper. A single cavity mode (red beam) coherently interacts with various species $m$ and $m'$ of collective spins (blue spheres) through phase shifted photons $e^{- i \phi_m} a$ and $e^{- i \phi_{m'}} a$ (red ellipsoids) with coupling strengths $g_m$ and $g_{m'}$. The cavity mode is coupled (dark red squiggles) to a bath of extra-cavity modes (pale red beams) leading to decay at the rate $\kappa$. This setup generically realizes effective nonreciprocal interactions between spins of different species $m$ and $m'$ (dashed blue lines). Similarly, the spins experience single-particle incoherent decay at the rate $\Gamma$.}
    \label{fig: cavity schematic}
\end{figure}

At a microscopic level, the interactions between particles are inherently reciprocal as a consequence of Newton's third law; therefore, nonreciprocal interactions must be cleverly engineered to emerge \cite{ivlev_statistical_2015} by allowing a system to interact reciprocally with a bath, then tracing out the bath degrees of freedom. The resulting system dynamics may be characterized by effective nonreciprocal interactions which generically break time-reversal ($\mathcal{T}$) symmetry. A schematic depiction of this recipe is shown in Fig.~\ref{fig: cavity schematic}. Although nonreciprocal interactions break time-reversal symmetry, the system may still separately possess generalized $\mathbb{Z}_2$ parity ($\mathcal{P}$) symmetries or a combined parity-time ($\mathcal{PT}$) symmetry.

In manybody systems, nonreciprocal interactions between constituent particles can give rise to nonreciprocal phase transitions \cite{fruchart_non-reciprocal_2021, weis_generalized_2025}.
% in which a static phase is induced to rotate by tuning the balance of reciprocal and nonreciprocal interactions, thus establishing a dynamical limit-cycle phase explicitly or spontaneously breaking $\mathcal{PT}$-symmetry.
Although not a necessary condition, nonreciprocal phase transitions are closely related to the appearance of exceptional points, constituting an ``ultimate, unavoidable form of nonreciprocity'' \cite{fruchart_non-reciprocal_2021}. Realizing nonreciprocal phase transitions in quantum systems is physically challenging since the constituent particles are more microscopic and, in a loose sense, closer to the fundamental constraint of reciprocity in their interactions. Moreover, the definition of time-reversal symmetry in open quantum systems is subtle \cite{prosen_pt-symmetric_2012, huber_emergence_2020} and the existence of exceptional points can change in different limits \cite{minganti_quantum_2019}. That being said, proposals have been introduced \cite{chiacchio_nonreciprocal_2023, hanai_photoinduced_2024, nadolny_nonreciprocal_2025} to realize nonreciprocal phase transitions in quantum systems building off of previous work on the engineering of nonreciprocal interactions, most notably in the context of the non-Hermitian skin-effect \cite{yao_edge_2018, zhang_review_2022}.

A paradigmatic quantum system with a $\mathbb{Z}_2$ parity symmetry is the Hepp-Lieb-Dicke model \cite{dicke_coherence_1954, hepp_equilibrium_1973, hepp_superradiant_1973} describing the electric-dipole interaction between spin-$1/2$ particles and a single cavity mode of the electric field. This model is closely related to the Ising model (with all-to-all couplings) after eliminating the cavity degree of freedom \cite{damanet_atom-only_2019, paz_driven-dissipative_2021}, perhaps the quintessential example of a model with a $\mathbb{Z}_2$ parity symmetry. In the context of the Dicke model, the symmetry-broken phase is called the superradiant phase (as opposed to the normal phase) and corresponds to a coherent, nonzero occupation of the cavity mode. A recent review of the Dicke model can be found in Ref.~\cite{kirton_introduction_2019} including both equilibrium and driven-dissipative variations. 
% For the rest of this paper, we focus on the dissipative Dicke model (without drive) and, therefore, drop the modifier. 

A versatile platform which can realize the Dicke model is an atomic Bose-Einstein condensate (BECs) trapped in an optical cavity \cite{domokos_collective_2002, black_observation_2003, baumann_dicke_2010, nagy_dicke-model_2010}, where the superradiant phase transition is characterized as a process of self-organization and is related to the so-called ``supersolid'' phase transition \cite{leonard_supersolid_2017}. One facet of the platform's versatility is the ability to introduce multiple spin species via spinor BECs. In this context, the Dicke model was recently extended to a nonreciprocal (or non-standard) two-species model \cite{landini_formation_2018} and beyond \cite{lyu_nonreciprocal_2025}, observed experimentally \cite{dogra_dissipation-induced_2019}, and shown to exhibit a nonreciprocal phase transition to a dynamical limit-cycle phase \cite{chiacchio_nonreciprocal_2023} (referred to generally as the ``dynamical phase'' throughout this work). While the existence of limit-cycle states or persistent oscillations has been observed in generalized Dicke models, for example, the states identified in Ref.~\cite{bhaseen_dynamics_2012} and Ref.~\cite{kirton_superradiant_2018} with reciprocal couplings only, the dynamical phase in the nonreciprocal Dicke model is induced by the nonreciprocity of the interactions.

In this work, we investigate the model from Ref.~\cite{chiacchio_nonreciprocal_2023} further, as shown schematically in Fig.~\ref{fig: cavity schematic}, deriving a Redfield master equation to describe the effective dynamics of the spin-only system, thus generalizing the analysis of Ref.~\cite{damanet_atom-only_2019} which considers a single-species reciprocal model. At the mean-field level--exact in the thermodynamic limit for the single-species Dicke model \cite{carollo_exactness_2021}--we compare the predictions of the full spin-cavity model, the spin-only model described by the Redfield master equation, and the spin-only model obtained from adiabatic elimination of the cavity mode. We find that the spin-only model described by the Redfield master equation not only provides a better quantitative match to the full spin-cavity model, perhaps expected, but also makes qualitatively different predictions from the adiabatic elimination model, principally in how it captures the effect of single-particle incoherent decay on the normal phase. Additionally, we look closely at the dynamical limit-cycle phase in the case of explicitly broken $\mathcal{PT}$-symmetry and uncover a region of phase-coexistence characterized by hysteresis terminating at a codimension-two exceptional point \cite{hanai_non-hermitian_2019}, consistent with the framework proposed in \cite{fruchart_non-reciprocal_2021} for a model with explicitly broken $\mathcal{PT}$-symmetry. Lastly, we push beyond the mean-field theory analyzing the spin-only model in the low-number limit, finding signatures of the behavior in the thermodynamic limit in the steady-state density matrix and the spectrum of the Liouvillian.

The paper is organized as follows. In Sec.~\ref{sec: description}, we introduce the nonreciprocal Dicke model and describe the procedure to obtain the effective spin-only model by integrating out the photonic degrees of freedom resulting in the Redfield master equation Eq.~\ref{eq: spin only adjoint Redfield} including non-secular terms. The relevant symmetries of the open system at the level of the Liouvillian are summarized in Table~\ref{tab: symmetries}, clarifying the realization (or not) of $\mathcal{PT}$-symmetry in various limits. In Sec.~\ref{sec: derive mean-field}, we treat the model introduced in Sec.~\ref{sec: description} at the mean-field level, exploring the phase diagram via the fixed-points, limit-cycles, and bifurcations of the corresponding nonlinear dynamical system. Our primary results for this section are summarized in Fig.~\ref{fig: linear-stability normal state} and Fig.~\ref{fig: dynamical state} and discussed in Sec.~\ref{sec: normal phases transitions} and Sec.~\ref{sec: dynamical phase}. We conclude with Sec.~\ref{sec: quantum treatment} which discusses the model in the low-number limit. The superradiant phase is analyzed in Fig.~\ref{fig: superradiant steady-states} while the dynamical limit-cycle phase is analyzed in Fig.~\ref{fig: liouvillian gap}. 

\section{Description of the model} \label{sec: description}

\subsection{Description of the closed system}

We study a quantum system consisting of three subsystems: spin (S), cavity (C), and bath (B) of extra-cavity modes. The total Hamiltonian has the general form 
\begin{equation}
    H = H_\text{S} + H_\text{SC} + H_\text{C} + H_\text{CB} + H_\text{B}
\end{equation}
where $H_\text{S}$, $H_\text{C}$, and $H_\text{B}$ are noninteracting terms corresponding to the spin, cavity, and bath subsystems, respectively, $H_\text{SC}$ is the interaction between the spins and the cavity, and $H_\text{CB}$ is the interaction between the cavity and the bath. 
% At this point, we assume that there is no direct interaction between the spins and the bath, although we will later phenomenologically introduce dissipative spin terms. 
The spin subsystem consists of spin-$1/2$ particles (two-level-systems) which are divided into distinguishable species $m$. 
% The spin-up and spin-down states are separated by the energy $\epsilon$. 
Motivated by the microscopic derivation of the model (see Appendix~\ref{app: microscopic parameters}) we take the species labels $m = - M, - M + 1, \ldots, M - 1, M$ where $M \geq 0$ is a fixed integer. While $2 M + 1$ is the total number of possible species, we let $M^\star$ denote the number of species with nonzero population $N_m$ ($M^\star \leq 2 M + 1$). The case $M = 0$ and $M^\star = 1$ corresponds to the single-species model, while the case $M = 1$ and $M^\star = 2$ corresponds to a two-species model and will be the focus of this work. 
% A spin $j$ of species $m$ is individually described by the vector spin-$1/2$ operator $\bm{s}^{(m, j)} = \left( s_1^{(m, j)}, s_2^{(m, j)}, s_3^{(m, j)} \right)$ with eigenvalues $\pm 1/2$ ($\hbar = 1$) for each component $i = 1$, $2$, or $3$. The cavity is described by a single bosonic mode of the electric field $a$ of energy $\omega_c$. Generically, we will assume that $\omega_c > 0$ microscopically corresponding to red-detuned pumping of the spins, see Appendix~\ref{app: microscopic parameters}, however interesting dynamical limit-cycle phases have also been investigated in the blue-detuned case \cite{kirton_superradiant_2018, kesler_emergent_2019}. The bath is described by a countable collection of bosonic extra-cavity modes of the electric field $b_k$ of energy $\xi_k$. 
The noninteracting terms in the Hamiltonian are therefore ($\hbar = 1$)
\begin{align}
    H_\text{S} & = \epsilon \sum_{m = - M}^M S_3^{(m)} \\
    H_\text{C} & = \omega_c a^\dagger a \\
    H_\text{B} & = \sum_k \xi_k b_k^\dagger b_k
\end{align}
where $\epsilon$ is the individual spin energy splitting,  $\bm{S}^{(m)} = \sum_{j = 1}^{N_m} \bm{s}^{(m, j)}$ is the vector collective spin-$\left(N_m / 2 \right)$ operator for the $N_m$ spin-$1/2$ particles of species $m$ constructed the individual spin-$1/2$ operators $\bm{s}^{(m, j)}$, $\omega_c$ is the cavity energy, and $a$ and $a^\dagger$ the annihilation and creation operators of the cavity mode, respectively. Generically, we will assume that $\omega_c > 0$ microscopically corresponding to red-detuned pumping (see Appendix~\ref{app: microscopic parameters}) however interesting dynamical limit-cycle phases have also been investigated in the blue-detuned case \cite{kirton_superradiant_2018, kesler_emergent_2019}. The interactions between the subsystems are described by the terms ($\hbar = 1$)
\begin{align}
    H_\text{SC} & = \sum_{m = - M}^M \frac{ 2 g_m }{ \sqrt{ N_m } } S_1^{(m)} \left( e^{- i \phi_m} a + e^{i \phi_m } a^\dagger \right) \\
    H_\text{CB} & = \sum_k h_k \left( a^\dagger b_k + a b_k^\dagger \right)
\end{align}
where we take the couplings $g_m$ and $h_k$ to be real. The normalizing factor of $\sqrt{N_m}$ is introduced to ensure extensive scaling in the thermodynamic limit. The spin-cavity interaction $H_\text{SC}$ includes counter-rotating terms essential to the Dicke model, while we assume only co-rotating terms are sufficient to describe the cavity-bath coupling in $H_\text{CB}$ (i.e. making the rotating-wave approximation).

We impose additional conditions on the coupling strengths $g_m$ and the phases $\phi_m$ which are motivated microscopically, see Appendix~\ref{app: microscopic parameters} for details. The coupling strengths are symmetric under swapping of species $m \leftrightarrow - m$, $g_m = g_{-m}$ while the phases are anti-symmetric, $\phi_m = - \phi_{-m}$, necessitating $\phi_0 = 0$. Importantly, for a multi-species model $M^\star > 1$, the phase shift of the cavity mode cannot, in general, be removed via gauge transformation of the cavity or spin operators \cite{chiacchio_dissipation-induced_2019}, a requirement for nonreciprocity \cite{clerk_introduction_2022}. 

\subsection{Description of the open system with the cavity}

The cavity-bath interaction $H_\text{CB}$ is used to impose dissipative dynamics on the spin-cavity subsystem. In the standard treatment of cavity dissipation, we trace out the bath modes assuming zero temperature and a (continuous) flat spectral density or, equivalently, delta-function time-correlations. This assumption is valid for many finely spaced extra-cavity modes spanning a large spectral range with constant coupling strengths $h_k = h$. Doing so, we obtain the Gorini–Kossakowski–Sudarshan–Lindblad (GKSL) master equation for the density matrix \cite{breuer_theory_2002} with quantum jump operator $a$ at the rate $\kappa$. The adjoint GKSL master equation for a generic spin or cavity operator $A$ follows as
\begin{equation} \label{eq: spin and cavity adjoint GKSL}
    \hat{\mathcal{L}}^* A = i \left[ H_\text{S} + H_\text{SC} + H_\text{C}, A \right] + \kappa \hat{\mathcal{D}}^* \left[ a \right] A + \Gamma \sum_{m = - M}^M \sum_{j = 1}^{N_m} \hat{\mathcal{D}}^* \left[ s_-^{(m, j)} \right] A
\end{equation}
where $\hat{\mathcal{L}}^*$ is the adjoint Liouvillian (in this case Lindbladian) superoperator and $\hat{\mathcal{D}}^* \left[ A \right] B = 2 A^\dagger B A - \left\{ A^\dagger A, B \right\}$ is the adjoint dissipation superoperator. We extract the time-evolution of correlation functions as $d \left\langle A \right\rangle / dt = \left\langle \hat{\mathcal{L}}^* A \right\rangle$ \cite{fazio_many-body_2025}. Here, we note that the Lamb shift is exactly zero in the limit of strict white-noise, while for a sharp energy cutoff it introduces a logarithmically-divergent correction to the bare cavity frequency, absorbed into $\omega_c$.

Single-particle incoherent decay is introduced phenomenologically with the quantum jump operator $s_-^{(m, j)} = s_1^{(m, j)} - i s_2^{(m, j)}$ at the rate $\Gamma$. This process arises from coupling the individual spins to an additional bath of bosonic phonon modes and formally yields a Lamb shift similar to that described above. For each variation of the model subsequently discussed, our treatment of single-particle incoherent decay is identical, with any renormalization corrections absorbed into $\epsilon$. This single-particle effect is not only essential to stabilizing the normal state when $\phi_m \neq 0$ \cite{buca_dissipation_2019, chiacchio_dissipation-induced_2019}, as will be discussed in more depth later, but is also experimentally relevant. Moreover, it has been shown that single-particle incoherent decay can counter-act dephasing effects \cite{kirton_suppressing_2017}, restoring the superradiant transition in a single-species model. On the other hand, it violates the conservation of total spin which renders theoretical approaches such as standard Holstein-Primakoff expansions \cite{kirton_introduction_2019, buca_dissipation_2019} and coherent-state path integral approaches \cite{torre_keldysh_2013} inapplicable, though  other field theory approaches have been studied \cite{torre_dicke_2016}. Collective incoherent decay with quantum jump operator $S_-^{(m)}$ is sometimes studied \cite{gelhausen_many-body_2017, gelhausen_dissipative_2018}, but, unlike single-particle incoherent decay, it does not stabilize the normal state.

Open quantum systems permit two distinct types of symmetries (which coincide in closed systems), weak (classical) and strong (quantum), distinguished by whether or not left and right symmetry-charges are separately conserved by the Liouvillian \cite{buca_note_2012, fazio_many-body_2025}.
% In the case of strong symmetries, there exists a steady-state solution for the density matrix in each symmetry sector, whereas for a weak symmetry, the density matrix still generically approaches a unique steady-state. Here, the open system possesses weak symmetries only. 
Although the continuous rotational symmetry of the spins has been broken, we introduce two important $\mathbb{Z}_2$ symmetries: (1) Superradiant parity (SR parity) symmetry with respect to the parity operator $\Pi = e^{i \pi \left( a^\dagger a + \sum_m S_3^{(m)} \right)}$. The SR parity symmetry arises due to parity conservation (in this case referring to even or odd) of the total excitation number, since the non-number-conserving counter-rotating terms in $H_\text{SC}$ can only create or annihilate excitations in pairs. (2) Nonreciprocal parity-time symmetry (NR $\mathcal{PT}$) with respect to the combined bipartite parity and time-reversal operator $\mathbb{PT}$ \cite{huber_emergence_2020}, where the unitary bipartite parity involution $\mathbb{P} \left( A \otimes B \right) \mathbb{P} = B \otimes A$ interchanges operators on the tensor product space $\mathcal{H} \otimes \mathcal{H}$ (with $\mathcal{H}$ a spin Hilbert spaces) while the anti-unitary time-reversal involution $\mathbb{T} A \mathbb{T} = A^\dagger$ performs Hermitian conjugation, thus interchanging pump and decay processes. We can also combine the operators $\mathbb{P}$ and $\mathbb{PT}$ with a gauge transformation of the spins or the cavity mode, which will be important in Sec.~\ref{sec: dynamical phase}. Lastly, since the spins are identically prepared and, in principal, not individual addressable, the system also has a weak permutation symmetry which formally arises from an $SU \left( 4 \right)$-symmetry of the Liouvillian superoperator \cite{xu_simulating_2013, hartmann_generalized_2016}. The symmetry is strong if there are no single-particle dissipative processes. The symmetries of Eq.~\ref{eq: spin and cavity adjoint GKSL} are summarized in Table~\ref{tab: symmetries}.

\subsection{Description of the open system without the cavity}

\begin{figure}[t]
    \centering
    \includegraphics[width=0.57\linewidth]{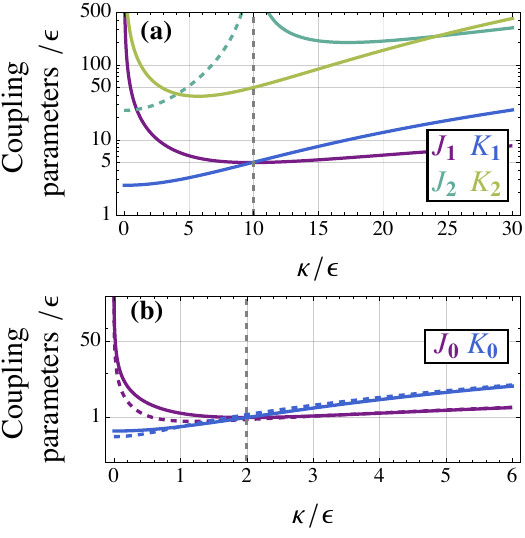}
    \caption{Effective coupling parameters in the spin-only system. (a) Coupling parameters $J_1 / \epsilon$, $K_1 / \epsilon$, $J_2 / \epsilon$, and $K_2 / \epsilon$ from Eq.~\ref{eq: spin only adjoint Redfield} as a function of the cavity decay rate $\kappa / \epsilon$ with the cavity energy $\omega_c / \epsilon = 10$ fixed (dashed gray line). $J_2$ is negative (dashed line plotting $- J_2$) for $\kappa^2 < \omega_c^2 - \epsilon^2$. (b) Coupling parameters $J_0 / \epsilon$ and $K_0 / \epsilon$ from Eq.~\ref{eq: spin only Redfield fast-cavity} as a function of the cavity decay rate $\kappa / \epsilon$ with the cavity energy $\omega_c = 2 \epsilon$ fixed (dashed gray line) compared to $J_1 / \epsilon$ and $K_1 / \epsilon$. There is no analogy to $J_2 / \epsilon$ and $K_2 / \epsilon$ in Eq.~\ref{eq: spin only Redfield fast-cavity}.
    % The coupling parameters $J_1 / \epsilon$ and $K_1 / \epsilon$ (dashed purple and blue respectively) are compared to $J_0 / \epsilon$ and $K_0 / \epsilon$ using a smaller value of $\omega_c / \epsilon$ in order to accentuate differences. There is no analogy to $J_2 / \epsilon$ and $K_2 / \epsilon$ in the system obtained from adiabatic elimination.
    } 
    \label{fig: coupling parameters}
\end{figure}

To obtain a description only in terms of the spins, we trace out the cavity mode in addition to the bath modes (still assuming zero temperature). The enlarged bath is now structured, and the previous assumption that the bath modes possess a flat spectral density is consistent with a Lorentzian spectral density of the cavity-bath eigenmodes. The time-correlations are, therefore, finite but exponentially suppressed \cite{damanet_atom-only_2019} with the form $e^{- i \Delta \phi} e^{- i \omega_c t - \kappa \left| t \right|}$ where $\Delta \phi = \phi_m - \phi_{m'}$ encodes information on the relative phase of the spin-cavity interaction. We assume that the inverse decay rate $1 / \kappa$ is sufficiently short compared to the time-scales of interest, allowing us to simplify the resulting master equation by extending the integration bound of the memory kernel, see Appendix~\ref{app: effective spin-only dynamics} for details. Following this prescription, the adjoint Redfield master equation for a generic spin operator $A$ is
\begin{equation} \label{eq: spin only adjoint Redfield}
    \begin{split}
    \hat{\mathcal{L}}^* A & = i \left[ H_\text{S}, A \right] + i \sum_{m, m' = - M}^M \frac{ g_m g_{m'} e^{- i \Delta \phi}}{ \sqrt{ N_m N_{m'}} } \sum_{j, k = 1}^2 H^\text{ind.}_{j k} \left[ \left( L_j^{(m)} \right)^\dagger L_k^{(m')}, A \right] \\
    & \quad + \sum_{m, m' = - M}^M \frac{ g_m g_{m'} e^{- i \Delta \phi}}{ \sqrt{N_m N_{m'}} } \sum_{j, k = 1}^2 D_{jk} \hat{\mathcal{D}}^* \left[ L^{(m')}_k, L^{(m)}_j \right] A + \Gamma \sum_{m = - M}^M \sum_{\ell = 1}^{N_m} \hat{\mathcal{D}}^* \left[ s_-^{(m, \ell)} \right] A
    \end{split}
\end{equation}
where, again, $\hat{\mathcal{L}}^*$ is the adjoint Liouvillian superoperator, $L^{(m)}_1 = S_-^{(m)}$ and $L^{(m)}_2 = S_+^{(m)}$ are the quantum jump operators for each species $m$. The coefficient matrix $H^\text{ind.}$ of the coherent bath-induced Hamiltonian is
\begin{subequations}
\begin{align}
     H^\text{ind.} & = - \frac{1}{4} \begin{bmatrix} \frac{ \omega_c - \epsilon}{ \kappa^2 + \left( \omega_c - \epsilon \right)^2 } & \frac{ \omega_c }{ \left( \kappa + i \epsilon \right)^2 + \omega_c^2 } \\ \frac{ \omega_c }{ \left( \kappa - i \epsilon \right)^2 + \omega_c^2 } & \frac{ \omega_c + \epsilon}{ \kappa^2 + \left( \omega_c + \epsilon \right)^2 } \end{bmatrix} \\
     & = - \frac{1}{16} \begin{bmatrix} \frac{1}{K_1} - \frac{1}{J_2} & \frac{1}{K_1} - \frac{i}{K_2} \\ \frac{1}{K_1} + \frac{i}{K_2} & \frac{1}{K_1} + \frac{1}{J_2} \end{bmatrix}
\end{align}
\end{subequations}
while the bath-induced dissipator is expanded in terms of the off-diagonal adjoint dissipation superoperator $\hat{\mathcal{D}}^* \left[ A, B \right] C = 2 B^\dagger C A - \left\{ B^\dagger A, C \right\}$ with rate-matrix
\begin{subequations}
\begin{align}
    D & = \frac{1}{4} \begin{bmatrix} \frac{ \kappa }{ \kappa^2 + \left( \omega_c - \epsilon \right)^2 } & \frac{ \kappa + i \epsilon }{ \left( \kappa + i \epsilon \right)^2 + \omega_c^2 } \\ \frac{ \kappa - i \epsilon }{ \left( \kappa - i \epsilon \right)^2 + \omega_c^2 } & \frac{ \kappa }{ \kappa^2 + \left( \omega_c + \epsilon \right)^2 } \end{bmatrix} \\
    & = \frac{1}{16} \begin{bmatrix} \frac{1}{J_1} + \frac{1}{K_2} & \frac{1}{J_1} - \frac{i}{J_2} \\ \frac{1}{J_1} + \frac{i}{J_2} & \frac{1}{J_1} - \frac{1}{K_2} \end{bmatrix}
\end{align}
\end{subequations}
Here, we have introduced the effective coupling parameters $J_1$, $J_2$, $K_1$, and $K_2$ which have units of energy and are given in terms of the system parameters $\kappa$, $\omega_c$, and $\epsilon$ as
\begin{subequations} \label{eq: coupling parameters}
\begin{align} 
    J_1 & = \frac{\Omega^4}{ 4 \kappa \left( \kappa^2 + \omega_c^2 + \epsilon^2 \right) } \\
    K_1 & = \frac{\Omega^4}{ 4 \omega_c \left( \kappa^2 + \omega_c^2 - \epsilon^2 \right) } \\
    J_2 & = \frac{\Omega^4}{ 4 \epsilon \left( \kappa^2 - \omega_c^2 + \epsilon^2 \right) } \\
    K_2 & = \frac{\Omega^4}{ 8 \kappa \omega_c \epsilon }
\end{align}
\end{subequations}
with the normalizing factor $\Omega^4 = \left( \kappa^2 + \omega_c^2 \right)^2 + 2 \left( \kappa^2 - \omega_c^2 \right) \epsilon^2 + \epsilon^4$. Both matrices $H^\text{ind.}$ and $D$ are then easily decomposed into a sum of Pauli matrices $\left( 1, \sigma_1, \sigma_2, \sigma_3 \right)$ as $H^\text{ind.} = \left( 1 / K_1 \right) + \left( 1 / K_1 \right) \sigma_1 + \left( 1 / K_2 \right) \sigma_2 - \left( 1 / J_2 \right) \sigma_3$ and $D = \left( 1 / J_1 \right) + \left( 1 / J_1 \right) \sigma_1 + \left( 1 / J_2 \right) \sigma_2 + \left( 1 / K_2 \right) \sigma_3$. The effective coupling parameters $K_1$ and $K_2$ are related to reciprocal interactions while $J_1$ and $J_2$ are related to nonreciprocal interactions. The time-evolution of correlation functions are similarly extracted as $d \left\langle A \right\rangle / dt = \left\langle \hat{\mathcal{L}}^* A \right\rangle$ \cite{fazio_many-body_2025}. We keep the bath-induced Hamiltonian which renormalizes the system Hamiltonian $H_S$, capturing polaritonic corrections to the bare spin energies, a practice which is not always necessary depending on the coupling strength and spectral width of the bath \cite{correa_potential_2024}, but is consistent with previous studies \cite{damanet_atom-only_2019}. An alternative form of Eq.~\ref{eq: spin only adjoint Redfield} is presented in Appendix~\ref{app: effective spin-only dynamics} (see Eq.~\ref{eq: spin only adjoint Redfield alt}) which makes clear the connection to the mean-field equations of motion, discussed later.

The strengths of the effective coupling parameters for a typical experimentally-realizable situation are shown in Fig.~\ref{fig: coupling parameters}a. For the purposes of studying the mediated interactions between the spins, the system is often tuned to the fast-cavity limit $\kappa, \omega_c \gg \epsilon$. In this regime, there is a special point $\kappa = \omega_c$ where the coupling parameters satisfy $J_1 \approx K_1$ and $\left| J_2 \right| \gg J_1$, $K_1$, and $K_2$ (so long as $\kappa = \omega_c$ is large enough), therefore resulting in coherent Hamiltonian interactions and dissipative interactions which are approximately equal in strength.

We stress that in deriving Eq.~\ref{eq: spin only adjoint Redfield} we have not made the secular approximation. Including nonsecular terms breaks the complete-positivity (but not the Hermitian-preservation) of Eq.~\ref{eq: spin only adjoint Redfield}, indicated by the lack of a positive-semidefinite rate-matrix $D$, and can lead to unphysical results for the density matrix or correlation functions for certain initial conditions, specifically in the transient behavior \cite{eastham_bath-induced_2016}. Notwithstanding this point, including nonsecular terms correctly describes the known behavior of open Dicke models, in particular the superradiant transition \cite{damanet_atom-only_2019}, as well as the behavior of other multi-component systems interpolating from diagonal to off-diagonal (coherence between decay and pumping) dissipative processes \cite{eastham_bath-induced_2016}. 

With the spin-only model Eq.~\ref{eq: spin only adjoint Redfield}, we can consider the SR parity and NR $\mathcal{PT}$ symmetries discussed in the previous section, making the small adjustment $\Pi = e^{i \pi \sum_m S_3^{(m)}}$ since we have integrated-out the cavity mode. The symmetries of Eq.~\ref{eq: spin only adjoint Redfield} are summarized in Table~\ref{tab: symmetries}. 

Despite nonsecular terms, Eq.~\ref{eq: spin only adjoint Redfield} can be brought into Lindblad form in the fast-cavity limit $\kappa, \omega_c \gg \epsilon$ \cite{damanet_atom-only_2019}
\begin{equation}  \label{eq: spin only Redfield fast-cavity}
    \begin{split}
    \hat{\mathcal{L}}^* A & = i \left[ H_\text{S}, A \right] - i \frac{1}{4 K_0} \sum_{m, m' = - M}^M \frac{ g_m g_{m'} e^{- i \Delta \phi}}{ \sqrt{ N_m N_{m'}} } \left[ S_1^{(m)} S_1^{(m')}, A \right] \\
    & \quad + \frac{1}{4 J_0} \sum_{m, m' = - M}^M \frac{ g_m g_{m'} e^{- i \Delta \phi}}{ \sqrt{N_m N_{m'}} } \hat{\mathcal{D}}^* \left[ S_1^{(m')}, S_1^{(m)} \right] A + \Gamma \sum_{m = - M}^M \sum_{\ell = 1}^{N_m} \hat{\mathcal{D}}^* \left[ s_-^{(m, \ell)} \right] A
    \end{split}
\end{equation}
where now $K_0 = \left( \kappa^2 + \omega_c^2 \right) / 4 \omega_c$ and $J_0 = \left( \kappa^2 + \omega_c^2 \right) / 4 \kappa$ are defined in analogy to the parameters $K_1$ and $J_1$; in particular, $K_1 \sim K_0$ and $J_1 \sim J_0$ in the fast-cavity limit $\kappa, \omega_c \gg \epsilon$, as shown in Fig.~\ref{fig: coupling parameters}b. Neglecting the energy splitting $\epsilon$ allows the system to sample the bath at the same frequency, and therefore the rate-matrix $D$ becomes constant. The symmetries of Eq.~\ref{eq: spin only Redfield fast-cavity} are summarized in Table~\ref{tab: symmetries}.

\begin{table}[t]
    \centering
    \begin{tabular}{l | c c}
    Symmetry & Eq.~\ref{eq: spin and cavity adjoint GKSL} and Eq.~\ref{eq: spin only adjoint Redfield} & Eq.~\ref{eq: spin only Redfield fast-cavity} \\
    \hline
    Permutation & Yes & Yes \\
    Permutation ($\Gamma = 0$) & Yes (strong) & Yes (strong) \\
    SR parity $\Pi$ & Yes & Yes \\
    Bipartite parity $\mathbb{P}$ & $\phi = \frac{k \pi}{2}$ & $\phi = \frac{k \pi}{2}$ \\
    $\mathbb{P}$ and gauge rotation & $\phi = \frac{\left( 2 k + 1 \right) \pi}{4} $ & $\phi = \frac{\left( 2 k + 1 \right) \pi}{4} $ \\
    Time-reversal $\mathbb{T}$ & No & No \\
    NR $\mathcal{PT}$ $\mathbb{P} \mathbb{T}$ & No & No \\
    $\mathbb{P} \mathbb{T}$ $\left( \Gamma = 0 \right)$ & No & Yes (strong)
    \end{tabular}
    \caption{Summary of the symmetries of the models considered (assuming a two-species model where appropriate): (Column 1) Spin and cavity Lindblad master equation Eq.~\ref{eq: spin and cavity adjoint GKSL} as well as the spin-only Redfield master equation Eq.~\ref{eq: spin only adjoint Redfield} including non-secular terms and (Column 2) spin-only Lindblad master equation in the fast-cavity limit $\kappa, \omega_c \gg \epsilon$ Eq.~\ref{eq: spin only Redfield fast-cavity} including non-secular terms. Symmetry operators are defined in the main text, and expressions for $\phi$ ($k \in \mathbb{Z}$) indicate where in parameter-space the symmetry is realized. The symmetry is weak, unless specified. 
    % We assume a two-species model $M = 1$ and $M^* = 2$ with balanced species populations $N_{-1} = N_1$ ($N_0 = 0$) and phase $\phi = \phi_1 = - \phi_{-1} = \Delta \phi / 2$.
    }
    \label{tab: symmetries}
\end{table}

\section{Analysis and results} \label{sec: analysis}

For the remainder of the paper, we focus on the spin-only system, although occasional comparison to the full spin-cavity system will prove helpful, for example, to benchmark results. The phase diagrams of single-species models have been investigated in much detail, including supplemental interaction terms and dissipative processes; some aspects of the single-species models generalize to the multi-species nonreciprocal model considered in this paper. In Sec.~\ref{sec: normal phases transitions} we study the so-called ``normal phase'' and explain how it loses stability at the mean-field level. In particular, we study the effect of single-particle incoherent decay, which we find is handled better by Eq.~\ref{eq: Redfield nonlinear mean-field EOM} derived from the Redfield master equation. In Sec.~\ref{sec: dynamical phase} we study the dynamical limit-cycle phase which is accessed from the normal phase via a supercritical Hopf bifurcation. We show that, in the case of explicitly broken $\mathcal{PT}$-symmetry, there is an additional phase transition within the predicted dynamical limit-cycle phase realized as a bifurcation of limit-cycles. Lastly, in Sec.~\ref{sec: quantum treatment}, we argue that prediction made at the mean-field level can be detected in systems with low particle numbers, supporting our analysis via exact numerical diagonalization exploiting the weak permutation symmetry. In the first section, Sec.~\ref{sec: derive mean-field}, we perform our analysis for a general multi-species model $M^\star > 2$. In subsequent sections, we focus on the two-species model and set the parameters $g = g_1 = g_{-1}$ and $\phi = \phi_1 = - \phi_{-1}$. 

% Overall, the dynamical system Eq.~\ref{eq: Redfield nonlinear mean-field EOM} produces a phase diagram in terms of the number of stable fixed-point solutions very similar to that presented in Ref.~\cite{chiacchio_nonreciprocal_2023}, employing adiabatic elimination as in Eq.~\ref{eq: AE mean-field EOM}. In the following section, however, we compare the predictions both qualitatively and quantitatively. Then we discuss details of the dynamical limit-cycle phase in Sec.~\ref{sec: dynamical phase}.

\subsection{Comparison of the mean-field equations of motion} \label{sec: derive mean-field}

We start by deriving the semiclassical equations of motion for the expectation values of the collective spin operators, comparing between those derived from the Redfield master equation Eq.~\ref{eq: spin only adjoint Redfield} and from semiclassical adiabatic elimination, equivalent at the mean-field level to using the Lindblad master equation in the fast-cavity limit $\kappa, \omega_c \gg \epsilon$ Eq.~\ref{eq: spin only Redfield fast-cavity} for the derivation. 

Consider Eq.~\ref{eq: spin only adjoint Redfield} first. We apply the adjoint Liouvillian superoperator to the collective spin operators $S_j^{(m)}$. To obtain the mean-field theory applicable in the thermodynamic limit $N_m \to \infty$, we take the expectation value and decouple all correlation functions, formally assuming that the second cumulants, equivalent to the second central moments, vanish. 
% This assumption is exact in the thermodynamic limit $N_m \to \infty$. 
We introduce the normalized scalar parameters $\tau_j^{(m)} = \left\langle S_j^{(m)} \right\rangle / N_m$ for each species $m$. The result is a closed system of equations in $3 \times \left( 2 M + 1 \right)$ variables, excluding the cavity field. We write the equations compactly in terms of the total vector $\bm{\tau} = \left( \left( \tau_1^{(m)} \right), \left( \tau_2^{(m)} \right), \left( \tau_3^{(m)} \right) \right)$ organizing all species $- M, \ldots, M$ into $x = 1$, $y = 2$, and $z = 3$ blocks. Then 
\begin{equation} \label{eq: Redfield nonlinear mean-field EOM}
    \dot{\bm{\tau}} = \begin{bmatrix} \bm{0}_v \\ \bm{0}_v \\ - \Gamma \bm{1}_v \end{bmatrix}  + \begin{bmatrix} - \Gamma \bm{I} & - \epsilon \bm{I} & \bm{0} \\ \epsilon \bm{I} + \bm{U} \left( \bm{\tau}_3 \right) & - \Gamma \bm{I} + \bm{V} \left( \bm{\tau}_3 \right) & \bm{0} \\ - \bm{U} \left( \bm{\tau}_2 \right) & - \bm{V} \left( \bm{\tau}_2 \right) & - 2 \Gamma \bm{I} \end{bmatrix} \bm{\tau}
\end{equation}
where $\bm{0}_v$ and $\bm{1}_v$ are $\left( 2 M + 1 \right)$-component vectors of all zeros and ones, respectively, and $\bm{0}$ and $\bm{I}$ are the $\left( 2 M + 1 \right) \times \left( 2 M + 1 \right)$ zero and identity matrix, respectively. The quadratic nonlinearities are encoded in the $\left( 2 M + 1 \right) \times \left( 2 M + 1 \right)$ matrix-valued functions $\bm{U}$ and $\bm{V}$ which mix the species $m$ and $m'$
\begin{gather}
    U_{m m'} \left( \bm{\tau}_j \right) = 2 \sqrt{ \frac{N_{m'}}{N_m} } g_m g_{m'} \left( \frac{ \cos \Delta \phi }{K_1} +\frac{ \sin \Delta \phi }{J_1} \right) \tau_j^{(m)} \\
    V_{m m'} \left( \bm{\tau}_j \right) = 2 \sqrt{ \frac{N_{m'}}{N_m} } g_m g_{m'} \left( \frac{ \cos \Delta \phi }{K_2} + \frac{ \sin \Delta \phi }{J_2} \right) \tau_j^{(m)}
\end{gather}
In the thermodynamic limit $N_m \to \infty$, we neglect sub-extensive corrections to the energy $\epsilon$ and the single-particle incoherent decay rate $\Gamma$ which are present in dynamical equations for finite-sized systems, see Appendix~\ref{app: effective spin-only dynamics} for details. The phase-dependent coefficients in $\bm{U}$ and $\bm{V}$ are separated into reciprocal parts, expressed in terms of $\cos \Delta \phi$, and nonreciprocal parts, expressed in terms of $\sin \Delta \phi$, with strengths characterized by the parameters $K_1$ and $K_2$ or $J_1$ and $J_2$, respectively. We stress that the coefficients $J_1$ and $J_2$ do not appear (or are infinite) in treatments of the reciprocal Dicke model \cite{damanet_atom-only_2019}. We will mostly be concerned with the fast-cavity limit $\kappa, \omega_c \gg \epsilon$ and, in this case, we notice $\kappa = \omega_c$ yields a special point where $J_2$ is exceedingly large, as shown in Fig.~\ref{fig: coupling parameters}a, therefore rendering the nonreciprocity described by $\bm{V}$ small.

The fixed-points $\dot{\bm{\tau}} = 0$ of Eq.~\ref{eq: Redfield nonlinear mean-field EOM} 
% correspond to the roots of a $3 \times \left( 2 M + 1 \right)$-order polynomial equation; although they 
are cumbersome to solve for in full generality, but two immediate conclusions can be drawn. First, the $x$ and $y$ components of the spins are related as $-\Gamma \tau_1^{(m)} = \epsilon \tau_2^{(m)}$. Therefore, all steady-state solutions must lie in a plane defined by the normal $\left( \cos \beta, \sin \beta, 0 \right)$ for each species with $\beta = \tan^{-1} \left( - \Gamma / \epsilon \right) + \pi / 2$. Second, the incoherent, spin-down ``normal state'' solution $\bm{\tau}_\text{ns} = \left( \bm{0}_v, \bm{0}_v, \left( - 1 / 2 \right) \bm{1}_v \right)$ exists for all values of the system parameters.

We study the stability of the normal state fixed-point by expanding Eq.~\ref{eq: Redfield nonlinear mean-field EOM} to linear order in the deviation $\delta \bm{\tau} = \bm{\tau} - \bm{\tau}_\text{ns}$ to obtain
\begin{equation} \label{eq: linearized Redfield EOM}
    \delta \dot{\bm{\tau}} = \begin{bmatrix} - \Gamma \bm{I} & - \epsilon \bm{I} & \bm{0} \\ \epsilon \bm{I} + \bm{U} \left( - 1 / 2 \right) & - \Gamma \bm{I} + \bm{V} \left( - 1 / 2 \right) & \bm{0} \\ \bm{0} & \bm{0} & - 2 \Gamma \bm{I} \end{bmatrix} \delta \bm{\tau}
\end{equation}
The equations for the $z$-components of each species $m$ decouple and are solved by simple exponential decay $\delta \tau_3^{(m)} \left( t \right) \propto e^{- 2 \Gamma t }$. They do not affect the stability of the normal state fixed-point and, therefore, we can exclude them from our analysis. 
% The eigenvalues of the reduced $2 \left( 2 M + 1 \right) \times 2 \left( 2 M + 1 \right)$ matrix are obtained from the characteristic polynomial, which can be further reduced to a $\left( 2 M + 1 \right)$-order polynomial by Schur's formula.

As promised, we compare the mean-field dynamical system Eq.~\ref{eq: Redfield nonlinear mean-field EOM} resulting from formally tracing out the cavity and extra-cavity modes with the system of equations obtained via adiabatic elimination (semiclassical or otherwise).
% of the spin-cavity system at the semiclassical level. The semiclassical adiabatic elimination procedure yields identical results to starting from the Lindblad master equation in the fast-cavity limit $\kappa, \omega_c \gg \epsilon$ Eq.~\ref{eq: spin only Redfield fast-cavity} at the mean-field level.
Consider Eq.~\ref{eq: spin and cavity adjoint GKSL}. We apply the adjoint Liouvillian superoperator to the collective spin operators $S_j^{(m)}$ as well as the cavity mode $a$. We obtain the mean-field dynamical system similarly to before, introducing an additional scalar parameter for the cavity mode $\alpha = \left\langle a \right\rangle$. Demanding that the cavity stay in the steady-state $\dot{\alpha} = 0$, we can express $\alpha$ in terms of spin parameters only with the equations of motion depending on the real combination
\begin{equation} \label{eq: elimination condition}
    e^{-i \phi_m} \alpha + e^{i \phi_m} \alpha^* = - \frac{1}{2} \sum_{m = - M}^M \frac{ g_m }{ \sqrt{N_m} } \left( \frac{ \cos \Delta \phi}{K_0} + \frac{ \sin \Delta \phi}{J_0}\right) \tau_1^{(m)}
\end{equation}
Substitution of Eq.~\ref{eq: elimination condition} leads to a system of equations similar to Eq.~\ref{eq: Redfield nonlinear mean-field EOM}
\begin{equation} \label{eq: AE mean-field EOM}
    \dot{\bm{\tau}} = \begin{bmatrix} \bm{0}_v \\ \bm{0}_v \\ - \Gamma \bm{1}_v \end{bmatrix} + \begin{bmatrix} - \Gamma \bm{I} & - \epsilon \bm{I} & \bm{0} \\ \epsilon \bm{I} + \bm{W} \left( \bm{\tau}_3 \right) & - \Gamma \bm{I} & \bm{0} \\ - \bm{W} \left( \bm{\tau}_2 \right) & \bm{0} & - 2 \Gamma \bm{I} \end{bmatrix} \bm{\tau}
\end{equation}
where the quadratic nonlinearities are now encoded in the $\left( 2 M + 1 \right) \times \left( 2 M + 1 \right)$ matrix-valued function $\bm{W}$
\begin{equation}
    W_{m m'} \left( \bm{\tau}_j \right) = 2 \sqrt{ \frac{N_{m'}}{N_m} } g_m g_{m'} \left( \frac{ \cos \Delta \phi}{K_0} + \frac{ \sin \Delta \phi}{J_0}\right) \tau_j^{(m)} 
\end{equation}
In particular, we highlight the absence of the $y$-$z$ quadratic nonlinearities in the $y$-equations and the $y$-$y$ quadratic nonlinearities in the $z$-equation, i.e. there is no analogy to the matrix-valued function $\bm{V}$. This result is consistent with the fast-cavity limit $\kappa, \omega_c \gg \epsilon$ of Eq.~\ref{eq: Redfield nonlinear mean-field EOM}.

Similar to Eq.~\ref{eq: linearized Redfield EOM}, the linearization of Eq.~\ref{eq: AE mean-field EOM} around the normal state is
\begin{equation} \label{eq: linearized AE EOM}
    \delta \dot{\bm{\tau}} = \begin{bmatrix} - \Gamma \bm{I} & - \epsilon \bm{I} & \bm{0} \\ \epsilon \bm{I} + \bm{W} \left( - 1 / 2 \right) & - \Gamma \bm{I} & \bm{0} \\ \bm{0} & \bm{0} & - 2 \Gamma \bm{I} \end{bmatrix} \delta \bm{\tau}
\end{equation}
The structural differences of Eq.~\ref{eq: linearized Redfield EOM} and Eq.~\ref{eq: linearized AE EOM} are not particularly profound in the single-species case; the linear stability analysis predicts the same stability and bifurcations, albeit at slightly shifted positions due to renormalization of the rates. However, in the following section, we show that the linear stability analysis Eq.~\ref{eq: linearized AE EOM} of the adiabatic elimination model Eq.~\ref{eq: AE mean-field EOM} fails to predict the correct stability of the normal state fixed-point in a multi-species model $M^\star > 0$ due to small but nonzero couplings between different species $m$ in Eq.~\ref{eq: Redfield nonlinear mean-field EOM} (i.e. $\bm{V}$) which are set exactly to zero

\subsection{The normal phase and its stability} \label{sec: normal phases transitions}

\begin{figure*}[th]
    \centering
    \includegraphics[width=0.9\linewidth]{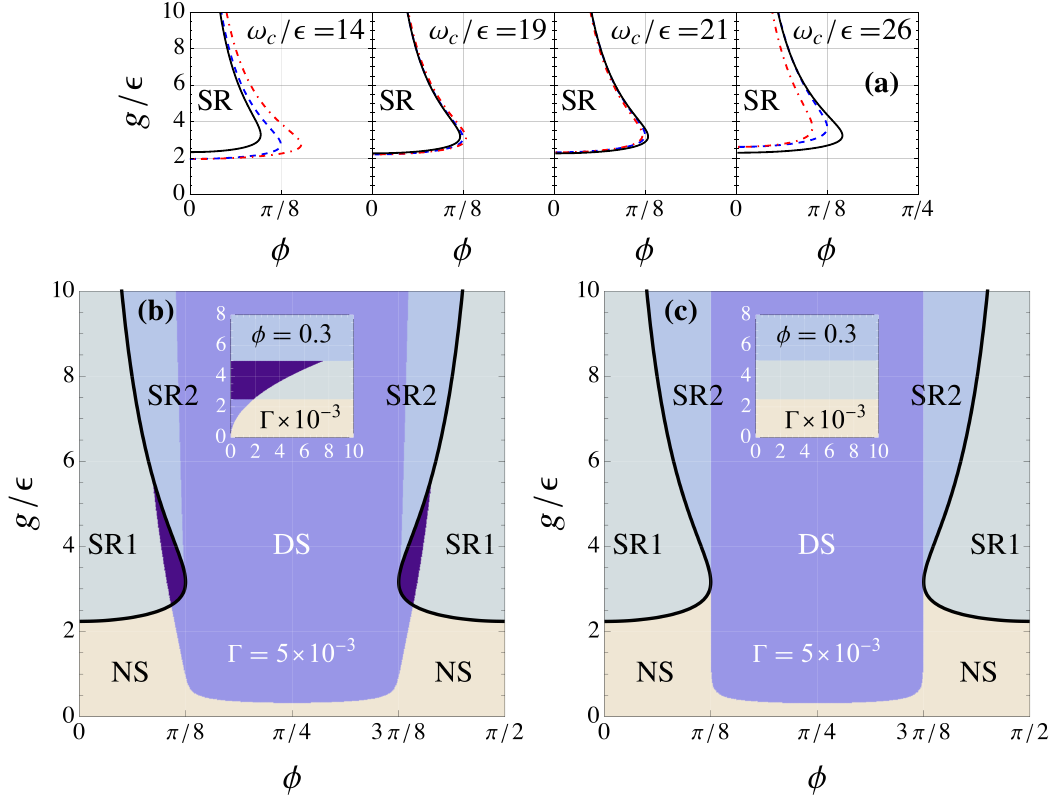}
    \caption{Comparison of the linear stability analyses of the normal state fixed-point $\bm{\tau}_\text{ns}$ for a two-species model. 
    (a) Boundary of the superradiant phase (lower branch, see text)
    % as indicated by the appearance of a pair of stable fixed-points with $\tau_3^{(m)} >0$ spontaneously breaking the SR parity symmetry (see text).
    for $\kappa / \epsilon = 20$, comparing the predictions from Eq.~\ref{eq: Redfield nonlinear mean-field EOM} (blue), Eq.~\ref{eq: AE mean-field EOM} (red), and Eq.~\ref{eq: spin and cavity adjoint GKSL} (black) which retains the cavity mode. 
    % For $\omega_c \approx \kappa$, the boundaries are nearly coincident, a special point discussed in the main text and also indicated in Fig.~\ref{fig: coupling parameters}a. We fix $\kappa / \epsilon = 20$;
    In the limit $\kappa / \epsilon \to \infty$, the region formally vanishes in the parameter $\phi$. 
    % (b, c) The linear stability analyses of the normal state fixed-point $\bm{\tau}_\text{ns}$ from Eq.~\ref{eq: linearized Redfield EOM} and Eq.~\ref{eq: linearized AE EOM} differ qualitatively, even with the cavity decay rate and cavity energy tuned to $\kappa / \epsilon = \omega_c / \epsilon = 20$. 
    (b) Linear stability analysis Eq.~\ref{eq: linearized Redfield EOM} and (c) linear stability analysis Eq.~\ref{eq: linearized AE EOM} for $\kappa = \omega_c = 20 \epsilon$. The main figures vary the phase $\phi$ and the coupling strength $g / \epsilon$ (fixing $\Gamma / \epsilon = 5 \times 10^{-3}$), while the insets vary the single-particle incoherent decay rate $\Gamma / \epsilon$ and the coupling strength $g / \epsilon$ (fixing $\phi = 0.3$). Color indicates the (in)stability of the normal state fixed-point $\bm{\tau}_\text{ns}$: Stable (``NS,'' tan), unstable with a pair of complex-conjugate eigenvalues (``DS,'' blue), unstable with one real eigenvalue (``SR1,'' green), unstable with two real eigenvalues (``SR2,'' light-blue), unstable with a real eigenvalue and a pair of complex-conjugate eigenvalues (unlabeled, dark-purple).
    }
    \label{fig: linear-stability normal state}
\end{figure*}

The normal phase is characterized by the stability of the normal state fixed-point $\bm{\tau}_\text{ns}$ as assessed by the eigenvalues of Eq.~\ref{eq: linearized Redfield EOM} or Eq.~\ref{eq: linearized AE EOM}. Loss of stability indicates a mean-field phase transition. In the two-species model of focus, we find that this can happen in three main ways: (1) Supercritical pitchfork bifurcation, (2) supercritical Hopf bifurcation, and (3) a combination of the two.

\subsubsection{Transition to the superradiant phase}

The steady-state superradiant phase transition results from a supercritical pitchfork bifurcation of the dynamical system Eq.~\ref{eq: Redfield nonlinear mean-field EOM} or Eq.~\ref{eq: AE mean-field EOM} and is marked by the determinant of the linear stability analyses Eq.~\ref{eq: linearized Redfield EOM} or Eq.~\ref{eq: linearized AE EOM} vanishing (for the quantum jump operators considered here \cite{kirton_suppressing_2017}), i.e. an eigenvalue becomes exactly zero. The critical coupling strength for the transition from the normal phase to the superradiant phase can be calculated analytically, above which there are two stable fixed-point solutions related by $\left( \tau_1^{(m)}, \tau_2^{(m)} \right) \leftrightarrow - \left( \tau_1^{(m)}, \tau_2^{(m)} \right)$ realizing the SR parity symmetry. In the single-species case, this reproduces the well-known critical coupling strength \cite{kirton_introduction_2019}
\begin{equation}
    g_c^{(1)}  \to \sqrt{ \frac{ \left( \kappa^2 + \omega_c^2 \right) \left( \Gamma^2 + \epsilon^2 \right) }{ 4 \omega_c\epsilon } }
\end{equation}
in the fast-cavity limit $\kappa, \omega_c \gg \epsilon$. In the two-species case, we find
\begin{equation} \label{eq: critical coupling Redfield}
    g_c^{(2)} = \sqrt{ \frac{ \Gamma^2 + \epsilon^2 }{ \left( \epsilon / K_1 - \Gamma / K_2 \right)} \frac{1}{f \left( \phi \right)} \left( 1 \pm \sqrt{1 - f \left( \phi \right)} \right) }
\end{equation}
where the dependence on the phase $\phi$ is described by the function
\begin{equation}
    f \left( \phi \right) = \left( 1 + \left( \frac{ \epsilon / J_1 - \Gamma / J_2}{ \epsilon / K_1 - \Gamma / K_2} \right)^2 \right) \sin^2 \left( 2 \phi \right)
\end{equation}
This result is consistent with the critical coupling strength originally reported in Ref.~\cite{landini_formation_2018}, expressed in terms of experimentally relevant parameters, up to renormalization corrections found here. The critical coupling strength $g_c^{(2)}$ may have two, one, or zero solutions, as shown in Fig.~\ref{fig: linear-stability normal state}. When it exists, the smaller solution (``$-$'', lower branch in Fig.~\ref{fig: linear-stability normal state}) indicates the transition from the normal phase to the superradiant phase. The larger solution (``$+$'', upper branch in Fig.~\ref{fig: linear-stability normal state}), though very close to indicating the emergence of new (unstable) fixed-point solutions, only provides local information on the stability of the normal state fixed-point $\bm{\tau}_\text{ns}$. In the fast-cavity limit $\kappa, \omega_c \gg \epsilon$, we can express the critical coupling strength for the two-species model $g_c^{(2)}$ in terms of the critical coupling strength for the single-species model $g_c^{(1)}$ presented earlier
\begin{equation} \label{eq: critical coupling Lindblad}
    g_c^{(2)} \to g_c^{(1)} \sqrt{ \frac{1}{\tilde{f} \left( \phi \right)} \left( 1 \pm \sqrt{ 1 - \tilde{f} \left( \phi \right)} \right)}
\end{equation}
where the function $f$ has been adjusted to $\tilde{f} \left( \phi \right) = \left( 1 + \left( \omega_c / \kappa \right)^2 \right) \sin^2 \left( 2 \phi \right)$. The expressions Eq.~\ref{eq: critical coupling Redfield} and Eq.~\ref{eq: critical coupling Lindblad} are compared in Fig.~\ref{fig: linear-stability normal state}a to the prediction following from Eq.~\ref{eq: spin and cavity adjoint GKSL} which retains the cavity mode. The expression Eq.~\ref{eq: critical coupling Redfield} is always closer to the prediction with the cavity mode, although there are significant deviations even in the fast-cavity limit $\kappa, \omega_c \gg \epsilon$. The expressions, however, are all very close for $\omega_c \approx \kappa$, the point discussed previously (and shown in Fig.~\ref{fig: coupling parameters}a) at which the coherent Hamiltonian interactions and dissipative interactions are approximately equal in strength. 

\subsubsection{Transition to the dynamical phase}

The transition to stable a limit-cycle state \cite{buca_dissipation_2019, chiacchio_dissipation-induced_2019, dogra_dissipation-induced_2019, chiacchio_nonreciprocal_2023} constituting the dynamical phase results from a supercritical Hopf bifurcation of the dynamical system Eq.~\ref{eq: Redfield nonlinear mean-field EOM} or Eq.~\ref{eq: AE mean-field EOM}. The features of the dynamical phase will be the focus of the following section; here, we focus on the transition from the normal phase. 

Even for $\omega_c = \kappa$ where the critical coupling strengths $g_c^{(2)}$ calculated from Eq.~\ref{eq: linearized Redfield EOM} and Eq.~\ref{eq: linearized AE EOM} quantitatively agree (see previous section), the location of the Hopf bifurcation may be significantly different, as shown in Fig.~\ref{fig: linear-stability normal state}b and c. In fact, in the limit of vanishing single-particle incoherent decay rate $\Gamma \to 0$, the linear instability of the normal state fixed-point $\bm{\tau}_\text{ns}$, as predicted by Eq.~\ref{eq: linearized Redfield EOM}, swells to wash-out the normal phase for all nonzero values of the coupling strength $g$ and phase $\phi$ \cite{buca_dissipation_2019}, as shown in the inset of Fig.~\ref{fig: linear-stability normal state}b for a cross-section of fixed $\phi$. This features is not predicted by Eq.~\ref{eq: linearized AE EOM} which, instead, predicts the persistence of the normal phase, as shown in the inset of Fig.~\ref{fig: linear-stability normal state}c for a cross-section of fixed $\phi$. The inclusion of single-particle incoherent decay, therefore, facilitates the transition to the dynamical phase at finite coupling strength $g$ and phase $\phi$.

The limit-cycle state grows continuously from the normal state fixed-point, as shown in Fig.~\ref{fig: dynamical state}a, where the time-averaged ``magnetization'' $\left\langle \tau_3^{(m)} \left( t \right) \right\rangle$ (top panel of Fig.~\ref{fig: dynamical state}a) as well as the time-averaged oscillation radius $\left\langle \sqrt{ \left( \tau_1^{(m)} \left( t \right) \right)^2 + \left( \tau_2^{(m)} \left( t \right) \right)^2 }  \right\rangle$ (middle panel of Fig.~\ref{fig: dynamical state}a) grow continuously from zero. As in the superradiant state, there are small quantitative differences in the limit-cycle states realized by Eq.~\ref{eq: Redfield nonlinear mean-field EOM} and Eq.~\ref{eq: AE mean-field EOM}. While the oscillation frequencies of the limit-cycle state (bottom panel of Fig.~\ref{fig: dynamical state}a) are nearly identical, discontinuously jumping to a nonzero value as a result of the Hopf bifurcation, the shapes of the limit-cycle states differ, as indicated by the differences in the time-averaged magnetization and oscillation radius. Though not shown, the differences are even clear for larger single-particle incoherent decay rates such that Eq.~\ref{eq: linearized Redfield EOM} and Eq.~\ref{eq: linearized AE EOM} make near identical predictions for the stability of the normal state fixed-point $\bm{\tau}_\text{ns}$. In any case, however, the limit-cycles are qualitatively very similar, as shown by the examples in the inset of Fig.~\ref{fig: dynamical state}a.

\subsubsection{More details on the stability of the normal phase}

Lastly, we note that Eq.~\ref{eq: linearized Redfield EOM} predicts the appearance of a codimension-two fold-Hopf bifurcation along the trajectory in Fig.~\ref{fig: linear-stability normal state}b from the normal phase to the dark-purple region, a feature which is not predicted by Eq.~\ref{eq: linearized AE EOM} as shown in Fig.~\ref{fig: linear-stability normal state}c. However, the system still flows to a stable superradiant fixed-point past the transition, and so this feature does not affect the long-time dynamics. 

For completeness, we also note that there exists regions of phase coexistence characterized by multiple linearly stable fixed-points with complementary basins of attraction \cite{keeling_collective_2010, soriente_dissipation-induced_2018}, consistent with the analysis in Ref.~\cite{chiacchio_nonreciprocal_2023}. This region, however, is not directly accessible from the normal phase. Moreover, it occurs at larger coupling strengths, potentially pushing beyond the validity of our perturbative analysis.

% The NR $\mathcal{PT}$-symmetry with respect to $\mathbb{P} \mathbb{T}'$ does not lead to a natural $\mathcal{PT}$-symmetry of the dynamical systems. Therefore, we interpret our results as well as those of Ref.~\cite{chiacchio_nonreciprocal_2023} as pertaining to explicitly broken NR $\mathcal{PT}$-symmetry at the mean-field level. 
% We note that, without single-particle incoherent decay $\Gamma = 0$, the NR $\mathcal{PT}$-symmetry with respect to $\mathbb{P} \mathbb{T}$ of the Liouvillian in Lindblad form for the spin-only model in the fast-cavity limit $\kappa, \omega_c \gg \epsilon$ \textit{does} yield a $\mathcal{PT}$-symmetry of the dynamical systems, namely $t \to - t$ and $\bm{\tau}_1 \to - \bm{\tau}_1$ or $\bm{\tau}_2 \to - \bm{\tau}_2$, flipping the chirality while leaving the quantization axis $\bm{\tau}_3$ invariant.

\subsection{The dynamical phase and consequences of parity and parity-time symmetry} \label{sec: dynamical phase}

\begin{figure*}[t]
    \centering
    \includegraphics[width = 0.93\linewidth]{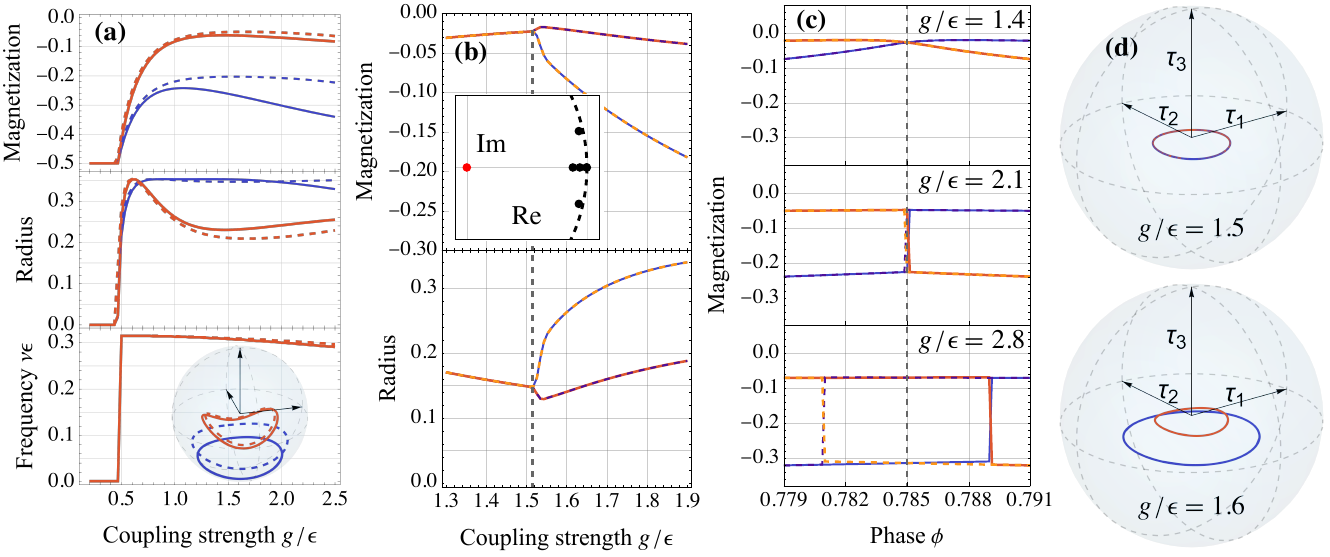}
    \caption{Analysis of the dynamical limit-cycle phase for a two-species model. (a) Behavior of the limit-cycle state for increasing coupling strength $g / \epsilon$ as characterized by the time-averaged magnetization, time-averaged oscillation radius, and the oscillation frequency for Eq.~\ref{eq: Redfield nonlinear mean-field EOM} (solid) and Eq.~\ref{eq: AE mean-field EOM} (dashed) and each species $m = + 1$ (blue) and $m = - 1$ (red). The phase $\phi = \pi / 4 - 0.1 \approx 0.68$ is slightly tuned away from $\phi = \pi / 4$ to differentiate the species (see b-d). (Inset) Example limit-cycle state in the Bloch sphere for $g / \epsilon = 2.8$.
    % using Eq.~\ref{eq: Redfield nonlinear mean-field EOM} (upper) and Eq.~\ref{eq: AE mean-field EOM} (lower) and separating each species, species $1$ ($m = + 1$, blue) and species $2$ ($m = - 1$, red). 
    (b) Time-averaged magnetization and time-averaged oscillation radius for each species $m = + 1$ (blue) and $m = - 1$ (red) and symmetric (solid) and anti-symmetric (dashed) initial perturbations as a function of the coupling strength $g / \epsilon$ 
    % within the dynamical limit-cycle phase 
    (using Eq.~\ref{eq: Redfield nonlinear mean-field EOM}). At $g / \epsilon \approx 1.51$, the system reaches a codimension-two exceptional point for $\Gamma \to 0$ and the 
    %formerly coincident 
    limit-cycles for each species differentiate. (Inset) Floquet multipliers inside the complex unit circle (dashed, center at red point) calculated as the eigenvalues of the monodromy matrix just below the transition. (c) Tuning the ``chirality breaking'' parameter $\phi$ uncovers a narrow region about $\phi = \pi / 4$ exhibiting hysteresis, growing wider for increasing coupling strength $g / \epsilon$. Symmetric (solid) and anti-symmetric (dashed) initial perturbations can converge to different stable limit-cycle states. (d) Example of the limit-cycle state in the Bloch sphere below the transition (upper) and above (lower). All plots use the parameters $\kappa / \epsilon = 20$ and $\omega_c / \epsilon = 20$ with (a-b) using $\Gamma / \epsilon = 1 \times 10^{-2}$ and (c-e) using $\Gamma / \epsilon = 5 \times 10^{-3} \epsilon$.
    }
    \label{fig: dynamical state}
\end{figure*}

As discussed in Sec.~\ref{sec: normal phases transitions}, perhaps the most striking feature of the model is the appearance of a dynamical limit-cycle phase. In this section, we look more closely at the features of the dynamical limit-cycle phase after the transition from the normal phase. We stress that Eq.~\ref{eq: spin only adjoint Redfield} and Eq.~\ref{eq: spin only Redfield fast-cavity} explicitly break the NR $\mathcal{PT}$ symmetry defined, see Table~\ref{tab: symmetries}, though the transition is still continuous \cite{hannukainen_dissipation-driven_2018}. Only in the fast-cavity limit $\kappa, \omega_c \gg \epsilon$ without single-particle incoherent decay $\Gamma = 0$ does the model have NR $\mathcal{PT}$ symmetry.

At the level of the dynamical systems Eq.~\ref{eq: Redfield nonlinear mean-field EOM} and Eq.~\ref{eq: AE mean-field EOM}, the unitary bipartite parity involution $\mathbb{P}$ is realized as $\bm{\tau}^{(- 1)} \leftrightarrow \bm{\tau}^{(1)}$. The dynamical systems are, then, left invariant if the additional transformation $\phi \to - \phi$ is made. Previous work \cite{chiacchio_nonreciprocal_2023} has used this combined transformation to define the NR $\mathcal{PT}$ symmetry at the mean-field level, interpreting $\phi \to - \phi$ as an effective time-reversal. Since this transformation involves the parameter $\phi$, however, the symmetric solution pairs generically exist at different points in parameter space. The consequence is that the transition from the normal phase to the dynamical phase (for $\phi \neq \pi / 4$) deterministically selects a $\mathcal{PT}$-broken limit-cycle state.
% as shown in Fig.~\ref{fig: dynamical state}a and b for $\phi = \pi / 4 - 0.1$. 
We note that in the fast-cavity limit $\kappa, \omega_c \gg \epsilon$ without single-particle incoherent decay $\Gamma = 0$, the existence of the NR $\mathcal{PT}$ symmetry results in a new symmetry of Eq.~\ref{eq: AE mean-field EOM} \cite{nakanishi_continuous_2025}, combining a standard time-reversal transformation $t \to - t$ and a spatial parity transformation $\bm{\tau}_1 \to - \bm{\tau}_1$, flipping the chirality while leaving the quantization axis $\bm{\tau}_3$ invariant (to be consistent with the SR parity symmetry, we could also take $\bm{\tau}_2 \to - \bm{\tau}_2$). 
% guarantees the existence of a NR $\mathcal{PT}$ symmetry of the mean-field dynamical system, using the results of Ref.~\cite{nakanishi_continuous_2025} under a Holstein-Primakoff approximation. Namely, 

Along the line $\phi = \pi / 4$, however, bipartite parity symmetry is restored up to a gauge transformation of the spins, for example $\left( \tau_1^{(1)}, \tau_2^{(1)} \right) \to - \left( \tau_1^{(1)}, \tau_2^{(1)} \right)$. This occurs, surprisingly, because the reciprocal interactions vanish. As shown in Fig.~\ref{fig: dynamical state}b and d (top row), the line $\phi = \pi / 4$ bisects regions with bipartite parity symmetry-broken limit-cycle states. As the strength of the nonreciprocity is increased via the coupling strength $g$ along the line $\phi = \pi / 4$, we find numerical evidence for a codimension-two exceptional point for $\Gamma \to 0$, as described in Ref.~\cite{fruchart_non-reciprocal_2021}, where the parity-symmetric limit-cycle state loses stability and is replaced by the coexistence of two stable parity-broken limit-cycle states. We identify the coexistence region by the appearance of hysteresis, as shown in Fig.~\ref{fig: dynamical state}c. In a small region of phases about $\phi = \pi / 4$, growing for increasing coupling strength $g$, whether the system is initialized with a symmetric (solid) or anti-symmetric (dashed) perturbation from the normal state fixed-point $\bm{\tau}_\text{ns}$ effects which stable, parity-broken limit-cycle state it converges to in the long-time limit. 

Lastly, as the coupling strength $g$ is increased even more near $\phi = \pi / 4$, we find dynamics which appear chaotic, as noted in Ref.~\cite{chiacchio_nonreciprocal_2023}, before the appearance of new stable fixed-points (the coexistence region described in Sec.~\ref{sec: normal phases transitions}) which ultimately eliminate the dynamical phase.

\subsection{Steady-states in the low-number limit} \label{sec: quantum treatment}

\begin{figure}[t]
    \centering
    \includegraphics[width=0.65\linewidth]{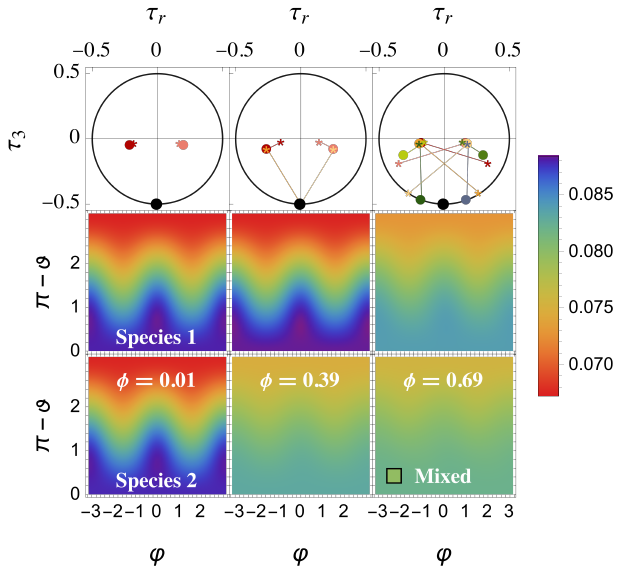}
    \caption{Superradiant steady-states in the fast-cavity limit $\kappa, \omega_c \gg \epsilon$ for phases $\phi = 0.01$, $\pi / 8 \approx 0.39$, and $\pi / 4 - 0.1 \approx 0.69$ and coupling strength $g / \epsilon = 8$. (Top row) Fixed-point solutions to the mean-field equations Eq.~\ref{eq: AE mean-field EOM} showing species $m = 1$ (filled circles) and $m = - 1$ (stars), connected by lines. There are three solutions (two stable) for $\phi = 0.01$, five solutions (two stable) for $\pi / 8$, and nine solutions (four stable) for $\pi / 4 -0.1$, distinguished by color, which all lie in a single plane (see text).
    % the plane defined by the normal $\left( \cos \beta, \sin \beta, 0 \right)$ with $\beta = \tan^{-1} \left( - \Gamma / \epsilon \right) + \pi / 2$ (see text). 
    (Middle and bottom row) For a finite-sized system (six particles each), we visualize the density matrix via the spin-averaged Wigner distribution $W^{(m)} \left( \vartheta, \varphi \right)$, species $m = + 1$ (middle row) and $m = - 1$ (bottom row), where $\vartheta$ is the polar angle and $\varphi$ the azimuthal angle, which is consistent with the mean-field predictions. The value is indicated by the color, with the completely mixed state $1/4 \pi \approx 0.08$ (green) for reference. 
    % For $\phi = 0.01$, the steady-states of each species are nearly identical, while for $\phi = \pi / 8$ species $m = - 1$ is more mixed. At $\phi = \pi / 4 - 0.1$ species $m = 1$ also becomes more mixed as bipartite parity symmetry is nearly restored. 
    We find a unique steady-state despite the change in the number of mean-field fixed-points. All plots use the parameters $\Gamma / \epsilon = 5 \times 10^{-3}$, $\kappa / \epsilon = 20$, and $\omega_c / \epsilon = 20$, as in Fig.~\ref{fig: liouvillian gap}.}
    \label{fig: superradiant steady-states}
\end{figure}

\begin{figure}[t]
    \centering
    \includegraphics[width=0.57\linewidth]{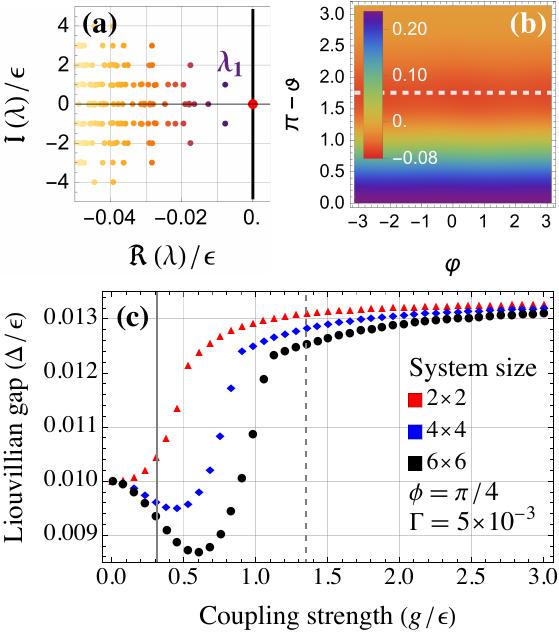}
    \caption{Signatures of the dynamical limit-cycle phase in a finite system in the fast-cavity limit $\kappa, \omega_c \gg \epsilon$. (a) Example spectrum of the Lindbladian Eq.~\ref{eq: spin only Redfield fast-cavity} near the steady-state eigenvalue $\lambda_0 = 0$ at coupling strength $g / \epsilon = 1.35$ and phase $\phi = \pi / 4$. (b) Spin-averaged Wigner distribution for species $m = + 1$ $W^{(1)} \left( \vartheta, \varphi \right)$, where $\vartheta$ is the polar angle and $\varphi$ the azimuthal angle, of the eigenmatrix corresponding to eigenvalue $\lambda_1$. There is a concentration of quasi-probability around the latitude $\pi - \vartheta$ of the mean-field limit-cycle (dashed gray line), comparable to the top panel of Fig.~\ref{fig: dynamical state}d. Both (a) and (b) use $6$ particles in each species and correspond to the dashed gray line in (c). (c) Liouvillian gap for various system sizes as a function of the coupling strength $g / \epsilon$ at fixed phase $\phi = \pi / 4$, 
    % At small coupling strength, the gap is set by a pair of spiraling complex conjugate eigenvalues with nonzero imaginary part; at larger coupling strengths, the gap is set by a decaying real eigenvalue. 
    decreasing towards the mean-field instability (gray solid line) as the system size increases. All plots use the parameters $\Gamma / \epsilon = 5 \times 10^{-3}$, $\kappa / \epsilon = 20$, and $\omega_c / \epsilon = 20$, as in Fig.~\ref{fig: superradiant steady-states}.}
    \label{fig: liouvillian gap}
\end{figure}

Despite the power of mean-field approaches applied to the Dicke model, away from the thermodynamic limit, the mean-field approach is expected to fail. In order to investigate systems with low particle numbers, we instead exactly solve the master equation $\dot{\rho} = \hat{\mathcal{L}} \rho$ numerically, an approach which is tractable due to the elimination of the cavity mode and the exploitation of permutation symmetry which both reduce the working dimension. Our goal is to explore how the mean-field predictions are realized at the level of the density matrix. In particular, we find (1) signatures a (nonreciprocal) superradiant state for $\phi \neq k \pi / 2$ shown in Fig.~\ref{fig: superradiant steady-states} and (2) signatures of the mean-field limit-cycles in the slowly-decaying eigenmodes of the Liouvillian $\mathcal{L}$ shown in Fig.~\ref{fig: liouvillian gap}. This analysis not only suggests the relevance of the mean-field predictions to smaller systems, but also highlights new features such as the appearance of completely mixed states. For the remainder of this section, we consider the spin-only model in the fast-cavity limit $\kappa, \omega_c \gg \epsilon$ Eq.~\ref{eq: spin only Redfield fast-cavity} in order to avoid possible numerical issues.  

Steady-states of the Liouvillian, which solve the simpler equation $\hat{\mathcal{L}} \rho_\text{ss} = 0$, define the phase of the system, similar to mean-field theory. First- and second-order phase transitions are, therefore, connected to non-analytic changes in $\rho_\text{ss}$ in the thermodynamic limit and are indicated by closing of the Liouvillian gap, defined later \cite{minganti_spectral_2018}. 

The dimension of the problem generically scales exponentially with the number of particles, $\mathcal{O} \left( 4^N \right)$ for a single species, thus rendering simple numerical approaches impracticable. However, the weak permutation symmetry indicated in Table~\ref{tab: symmetries} can be exploited to obtain polynomial scaling. To do so, we decompose the density matrix as 
\begin{equation}
    \rho = \sum_{\left( s_{-M}, \ldots, s_{M} \right)} p_{\left( s_{-M}, \ldots, s_{M} \right)} \rho_{\left( s_{-M}, \ldots, s_{M} \right)}
\end{equation}
where $p_{\left( s_{-M}, \ldots, \sigma_{M} \right)}$ is the probability of density matrix block $\rho_{\left( s_{-M}, \ldots, s_{M} \right)}$, and $s_m$ run over the allowed spins of each species $m$ (not to be confused with the individual spin operators $\bm{s}^{(m, j)})$, slightly generalizing the approach in Ref.~\cite{shammah_open_2018}. That is, there is no coherence between different spin manifolds $\left( s_{-M}, \ldots, s_{M} \right)$, even though single-particle dissipative processes break the total spin conservation. 
% While the dimension of the Hilbert space naively scales exponentially with particle number, $\mathcal{O} \left( 4^N \right)$ for a single species, permutation symmetry (and the assumption of a permutation-symmetric initial condition) restricts the dynamics of the density matrix to the symmetric subspace of the Hilbert space with a dimension that scales polynomially with particle number, $\mathcal{O} \left( N^3 \right)$ or even $\mathcal{O} \left( N^2 \right)$ for a single species when only diagonal processes (decay and pumping with no coherence) are included \cite{chase_collective_2008}. The reduction in the dimension leads to speed-up of numerical calculations such as finding the spectra, steady-states, or complete time-resolved dynamics. For multi-species models, however, the scaling is less powerful, going as $\mathcal{O} \left( N^{3 M^\star} \right)$ in general, i.e. $\mathcal{O} \left( N^6 \right)$ for the two-species model considered here. 
In practice, this allows us to simulate two-species systems with up to seven particles in each species,
%(roughly the equivalent to a single-species system with $42$ particles), 
enough to resolve distinct features of the steady-states and phases but not enough to investigate finite-size scaling and the connection to the classical limit.
To help visualize the density matrix, we first calculate the reduced density matrix for each species $\rho^{(m)}$, then calculate the Wigner (quasi-probability) distribution $W^{(m)}_{s_m} \left( \vartheta, \varphi \right)$ defined on the sphere parameterized by the polar angle $\vartheta$ and the azimuthal angle $\varphi$ \cite{klimov_group-theoretical_2009} for each allowed value of the spin $s_m$. Lastly, we calculate the spin-average Wigner distribution $W^{(m)} \left( \vartheta, \varphi \right)$, see Appendix~\ref{app: exact numerical diagonalization} for details.

\subsubsection{Superradiant steady-state}

For a single-species model, the steady-state is the completely mixed state $\rho \propto 1$ as a result of Hermitian quantum jump operators \cite{damanet_atom-only_2019}. The multi-species model still hosts nontrivial steady-states, however, which we explore in this section. 

We analyze the effect of nonreciprocity characterized by the phase $\phi$ on the superradiant state in the two-species model with $6$ particles in each species. The results are shown in Fig.~\ref{fig: superradiant steady-states} and compared to mean-field fixed-points. We use a larger value of the coupling strength $g$ to make superradiant signatures more apparent. For $\phi \approx 0$, the steady-states of species $m = 1$ and $m = - 1$ are nearly identical. At the mean-field level, there exists two fixed-point attractors with nonzero values of $\tau^{(m)}_3$ respecting the SR parity symmetry, and the locations of the fixed-point attractors in the coordinates of the two species are approximately equivalent (see the overlapping filled circles and stars). In the quantum system, the spin-averaged Wigner distribution exhibits resolved peaks at $\varphi = \tan^{-1} \left( - \Gamma / \epsilon \right)$ and $\tan^{-1} \left( - \Gamma / \epsilon \right) + \pi$, very close to $\varphi = 0$ and $\pi$ since $\Gamma \ll \epsilon$, and with latitude comparable to the mean-field prediction. The small system size reduces the contrast of the superradiant state from the normal state centered at the south pole $\vartheta = \pi$ (not shown). As the nonreciprocity is increased via the phase $\phi$, the locations of the fixed-point attractors in the coordinates of the two species differentiate, with species $m = 1$ moving away from the origin and species $m = - 1$ moving towards the origin. In turn, the peaks in the spin-averaged Wigner distribution of species $m = 1$ increase in contrast slightly, while the peaks in the spin-averaged Wigner distribution of species $m = - 1$ decreases in contrast approaching the constant value $W^{(m)} \left( \vartheta, \varphi \right) \approx 1 / 4 \pi$ associated with the completely mixed state $\rho^{(m)} \propto 1^{(m)}$, but not fully losing resolution of the superradiant peaks. For $\phi \to - \phi$, we find gain in contrast in species $m = - 1$ and loss of contrast in species $m = 1$, as expected by bipartite parity transformation. Finally, as the phase $\phi \to \pi / 4$, we observe the peaks of species $m = 1$ also decreases in contrast as bipartite parity symmetry is restored. At the mean-field level, multiple stable fixed-points coexist.

\subsubsection{Limit-cycle steady-state}

We identify signatures of the dynamical phase in the quantum system by analyzing the Liouvillian spectrum, eigenmatrices, and Liouvillian gap. In Fig.~\ref{fig: liouvillian gap}a, an example of the Liouvillian spectrum within the mean-field dynamical limit-cycle phase is shown for a system with $6$ particles in each species. The zero-eigenvalue $\lambda_0 = 0$ corresponds to the steady-state density matrix. The eigenvalue with the smallest nonzero real-part in magnitude is $\lambda_1$ ($\lambda_1^*$); the corresponding eigenmatrix decays at the slowest possible rate $\Re \left( \lambda_1 \right)$ and induces coherences which oscillate at the frequency $\Im \left( \lambda_1 \right)$. The Liouvillian gap $\Delta$ is, thus, naturally defined as $\Delta = - \Re \left( \lambda_1 \right)$ \cite{minganti_spectral_2018}. The appearance of limit-cycle states in the thermodynamic limit correspond to the simultaneous closing of the Liouvillian gap by a set of eigenvalues coming in complex conjugate pairs and equally spaced in their imaginary-part \cite{dutta_quantum_2025}. While large systems sizes are currently inaccessible in our approach, we investigate the Liouvillian gap $\Delta$ as a function of the coupling strength $g$ near the mean-field phase transition from the normal phase to the dynamical limit-cycle phase for $2$ and up to $6$ particles in each species, as shown in Fig.~\ref{fig: liouvillian gap}c setting the phase $\phi = \pi / 4$. At small coupling strength, the gap is set by a pair of spiraling complex conjugate eigenvalues with nonzero imaginary part, while at larger coupling strengths, the gap is set by a decaying real eigenvalue. The Liouvillian gap begins to decrease as the system size increases, close to the mean-field Hopf bifurcation in the thermodynamic limit (solid gray line in Fig.~\ref{fig: liouvillian gap}c). However, as shown in Fig.~\ref{fig: liouvillian gap}a, it is difficult to ascertain if the eigenvalues with larger imaginary part are approaching the imaginary axis with the proposed parabolic envelope \cite{dutta_quantum_2025}.

Lastly, the spin-averaged Wigner distribution of the reduced eigenmatrix for species $1$ ($m = +1$) corresponding to eigenvalue $\lambda_1$ is shown in Fig.~\ref{fig: liouvillian gap}b. We find a concentration of quasi-probability around the latitude of the mean-field limit-cycle state shown. We attribute the large width of the band is due to the small system size. Nevertheless, we argue that this result is consistent with the mean-field prediction.

\section{Conclusion}

In this work, we have developed and analyzed an effective spin-only description of the multi-species nonreciprocal Dicke model based on the Redfield master equation which improves upon adiabatic elimination approaches. Through a mean-field analysis, we have demonstrated that this approach not only provides better quantitative agreement with the full spin-cavity model but also captures crucial qualitative features. Additionally, in the case of explicitly broken $\mathcal{PT}$-symmetry, we have uncovered a region of phase coexistence between parity symmetry-broken limit-cycle states terminating at a codimension-two exceptional point. Furthermore, eliminating the cavity mode and leveraging the model's underlying weak permutation symmetry enabled us to probe the exact steady-states of the density matrix for system sizes larger than would otherwise be feasible. This analysis revealed that signatures of macroscopic mean-field phases, including both the superradiant phase and the onset of dynamical instabilities leading to limit-cycles states, are detectable in small systems. In future work, the fine-grained phases within the region of phase coexistence could be explored, as well as the transition to chaos. Additionally, expanding the accessible system size could help connect the quantum and classical limits in an important multi-species spin model.

\section{Acknowledgments}
The authors acknowledge useful discussions with Jonathan Keeling. JJ acknowledges useful discussion with Matteo Brunelli, Tobias Nadolny, and Michel Fruchart.

\paragraph{Funding information}
PBL is partially supported by the Moore Foundation grant GBMF12763. This research benefited from the Physics Frontier Center for Living Systems funded by the National Science Foundation (PHY-2317138).

\begin{appendix}
\numberwithin{equation}{section}

\section{Microscopic parameters} \label{app: microscopic parameters}

In this section of the appendix, we describe the experimental system motivating the model described in the main text.

We consider a system of atoms of mass $m_a$ interacting with high-frequency electric fields and a uniform bias magnetic field. In particular, the electric field is comprised of two components: a monochromatic classical ``pump'' field of frequency $\omega_p$ and a monochromatic quantum field of frequency $\omega_c'$ (different from the $\omega_c$ in the main text). The pump field consists of a single macroscopically occupied spatial-polarization mode labeled by the wavevector $\bm{k}_p$ and polarization $\bm{\varepsilon}_p$. The quantum field consists of a single spatial mode labeled by the wavevector $\bm{k}_c$ and two orthonormal polarization modes $\bm{\varepsilon}_c^{(j)}$ for $j = 1, 2$, since we have not assumed either to be macroscopically occupied. Finally, the classical pump field and quantum field are retro-reflected by mirrors to form standing waves termed the optical lattice and the cavity mode, respectively. The optical lattice has Bloch eigenstates $\left| n, q \right\rangle$ where $n$ is the band-index and $q$ is the quasimomentum. For simplicity, we choose the optical lattice to be aligned with the $z$-axis, $\bm{k}_p \propto \hat{\bm{z}}$. 

If we assume that the electric field is far-detuned from any electronic (optical) transitions in the atoms and that the intensity is sufficiently low to limit saturation effects, then the atom-field interaction is effectively governed by the dynamics of the magnetic sublevels coupled to the polarization modes of the field and is modeled by the interaction \cite{happer_optical_1972}
\begin{equation}
    V_d = \sum_{j,k} E_j^{(+)} \alpha_{jk} E_k^{(-)}
\end{equation}
where $E_j^{(\pm)}$ denotes the right-handed ($+$) and left-handed ($-$) frequency components of the $j$-th spatial component of the electric field $\bm{E}$. The subscript $d$ indicates that the interaction is derived in the electric dipole approximation. The operator $\alpha_{jk}$ is known as the polarizability tensor and can be separated into scalar, vector, and tensor parts
\begin{align}
    \alpha_{jk} & = \alpha^{(s)}_{jk} + \alpha^{(v)}_{jk} + \alpha^{(t)}_{jk} \\
    \alpha^{(s)}_{jk} & = \alpha_s \delta_{jk} \\
    \alpha^{(v)}_{jk} & = - i \frac{\alpha_v}{2f} \sum_\ell \epsilon_{jk\ell} f_\ell \\
    \alpha^{(t)}_{jk} & = \frac{\alpha_t}{2 f \left( 2 f - 1 \right)} \left( 3 \left( f_j f_k + f_k f_j \right) - 2 \bm{f}^2 \delta_{jk} \right) 
\end{align}
where the coefficients $\alpha_s$, $\alpha_v$, and $\alpha_t$ are the conventional scalar, vector, and tensor polarizabilities defined in Ref.~\cite{le_kien_dynamical_2013}, $f$ is the hyperfine angular momentum of the atoms in the electronic ground state, and $f_\ell$ are the associated $x$, $y$, and $z$ components of the spin-$f$ operators. In the main text, we set $M = f$. The polarizabilities are strong functions of frequency and exhibit resonance behavior. Therefore, although the hierarchy $\left| \alpha_s \right| > \left| \alpha_v \right| > \left| \alpha_t \right|$ may seem natural, it does not always hold. For example, in Cs, the scalar polarizability vanishes at $\lambda = 880.2$~nm between the D1 and D2 transitions while the vector polarizability remains appreciable \cite{le_kien_dynamical_2013}. This behavior may be an asset for experiments rather than a liability, allowing for dramatic tuning of the ratio $\alpha_v / \alpha_s$ which is a primary parameter in the Dicke model derived later. 

We consider the case of bosonic atoms ($f$ even) at low temperature, and assume the atoms form a Bose-Einstein condensate (BEC) in the zero-quasimomentum state $q = 0$ of the ground band $n = 0$ of the optical lattice. However, the optical lattice potential couples the BEC to the edges of the first Brillouin zone via Bragg transitions. Imposing orthogonality of the optical latices and cavity mode $\bm{k}_p \cdot \bm{k}_c = 0$ and neglecting transitions between the ground and excited bands assuming a large enough intensity of the pump field, we can construct a two-mode model in momentum-space. Moreover, if we assume a large enough uniform bias magnetic field $\left| \bm{B} \right|$ in the direction of the pump field $\bm{B} \parallel \bm{k}_p$, then we can neglect spin-exchange processes. This assumption ensures that the model is diagonal in the magnetic quantum number $m$ and that only one polarization mode of the cavity $\bm{\varepsilon}_c^{(1)}$ is coupled to the BEC. 

Overall, we obtain the Dicke model presented in the main text in the rotating frame of the pump field with the parameters given microscopically as
\begin{align}
    \epsilon & = E_L \left( k_p \right) - E_L \left( 0 \right) + \frac{k_c^2}{2m_a} - \omega_p \\
    \omega_c & = \omega_c' - \omega_p + 2 \left| E_c \right|^2 N_m \sum_{m = -f}^f \left( \alpha_s + \frac{\alpha_t \left( f \left( f + 1 \right) - 3 m^2 \right)}{2 f \left( 2 f - 1 \right)} \right) \\
    G_m & = 2 \sqrt{2} \nu E_p E_c\left( \alpha_s \cos \left( \varphi \right) - i \frac{\alpha_v}{2 f} \sin \left( \varphi \right) e^{- i \chi} m + \frac{\alpha_t \left( f \left( f + 1 \right) - 3 m^2 \right)}{2 f \left( 2 f - 1 \right)} \cos \left( \varphi \right) \right)
\end{align}
The energy splitting $\epsilon$ depends on the difference in eigenenergies between the Bloch state at the edge of the Brillouin zone $E_L \left( k_p \right)$ and the zero-quasimomentum state $E_L \left( 0 \right)$, the recoil energy of the cavity mode $k_c^2 / 2 m_a$, and and the energy of the pump field $\omega_p = k_p c$. The energy of the cavity mode $\omega_c$ is mostly given by the bare energy of the cavity mode $\omega_c'$ relative to the energy of the pump field $\omega_p$, but also includes a correction due to the presence of the BEC with strength proportional to the atom number $N_m$, where $E_c = \sqrt{\hbar \omega_c' / 2 \epsilon_0 V}$ is the amplitude normalization of the quantized electromagnetic field ($\epsilon_0$ is the permittivity of free space and $V$ is the quantization volume). Finally, $G_m$ is the complex coupling of the two-level systems in momentum-space to the cavity mode. The coupling depends on the macroscopic amplitude of the pump field $E_p$ (which we have chosen to be real by imposing a phase reference), the  amplitude normalization of the quantized electromagnetic field $E_c$, as well as an overlap integral $\nu$ between the potential term $\cos \left( k_p z \right)$, the Bloch state at the edge of the Brillouin zone $\left\langle 0, k_p \right|$, and the zero-quasimomentum state $\left| 0, 0 \right\rangle$. The polarization of the pump field is described by the parameters $\varphi$ and $\chi$ as $\bm{\varepsilon}_p = \left( \sin \left( \varphi \right) e^{i \chi}, \cos \left( \varphi \right), 0 \right)$. If we assume without loss of generality that the cavity mode is aligned with the $x$-axis, $\bm{k}_c \propto \hat{\bm{x}}$, then $\varphi$ gives the angle of linear polarization relative to the $y$-axis, normal to the cavity mode. Additionally, $\chi$ controls the degree of elliptic polarization, with $\chi = 0$ yielding linear polarization. The coupling strength $g_m$ and phase $\phi_m$ discussed in the main text are obtained as $g_m = \left| G_m \right|$ and $\phi_m = - \text{arg} \left( G_m \right)$. 

Consider the case of vanishing tensor-polarizability $\alpha_t = 0$ and linear polarization of the pump field $\chi = 0$. Then the effective Dicke model is characterized by the coupling strengths
\begin{equation}
    g_m = 2 E_p E_c \sqrt{\alpha_s^2 \cos^2 \left( \varphi \right) + \frac{\alpha_v^2 m^2}{4 f^2} \sin^2 \left( \varphi \right) }
\end{equation}
and the phases 
\begin{equation}
    \phi_m = - \tan^{-1} \left\{ \frac{ \alpha_v }{ \alpha_s } \frac{ m }{ 2 f } \tan \left( \varphi \right) \right\}
\end{equation}
Since $m$ only appears quadratically in $g_m$, it is clear that the coupling strengths are symmetric under swapping of species $m \leftrightarrow - m$, $g_m = g_{-m}$. Moreover, since inverse tangent is an odd function, it is also clear that the phases are anti-symmetric under swapping of species $m \leftrightarrow - m$, $\phi_m = - \phi_{-m}$. These are the conditions assumed in the main text.

All other parameters being equal, species with nonzero $m$ interact with the cavity mode with an enhanced strength
\begin{equation}
    g_m = g_0 \sqrt{ 1 + \left( \frac{\alpha_v}{\alpha_s} \right)^2 \frac{m^2}{4 f^2} \tan^2 \left( \varphi \right) }
\end{equation}
where $g_0 = 2 E_p E_c \left| \alpha_s \cos \left( \varphi \right) \right|$. In particular, the $m = 0$ species becomes ``dark'' when the pump and cavity polarization are aligned $\varphi = \pi/2$. Physically, this is because the when $\varphi = \pi/2$ then only the vector and tensor polarizability contribute to the interaction, which are zero for the $m = 0$ species. 
% For a fixed $g_0$, the coupling strengths $g_m$ are determined by the phases $\phi_m$ as $g_m = g_0 \sqrt{ 1 + \tan^2 \left( \phi_m \right) }$. 

In the main text, we primarily study the phase diagram in the theoretically-helpful parameters $\left( g_m, \phi_m \right)$. However, we can also parameterize the coupling strengths $g_m$ and phases $\phi_m$ as
\begin{align}
    g_m & = A_1 \left| \cos \left( \varphi \right) \right| \sqrt{ 1 + A_2^2 m^2 \tan^2 \left( \varphi \right) } \\
    \phi_m & = - \tan^{-1} \left\{ A_2 m \tan \left( \varphi \right) \right\}
\end{align}
where $A_1 = 2 E_p E_c \left| \alpha_s \right|$ and $A_2 = \alpha_v / 2 f \alpha_s$. Then we can study the distorted phase diagram in the parameters $A_1$, controlled by the intensity of the pump field, and $\varphi$, controlled by the pump field polarization, for fixed $A_2$, controlled by the ratio of vector and scalar polarizabilities.

\section{Derivation of the effective spin-only dynamics} \label{app: effective spin-only dynamics}

Integrating out the cavity mode in addition to the extra-cavity bath modes, we obtain the Redfield master equation describing the effective dynamics of the spin-only system. The standard procedure is outlined in the following section. 

Let $H_\text{SB}$ be the interaction between the system and the bath in the Schrödinger picture. In the main text, this is the interaction between the spins and the cavity $H_\text{SC}$. We consider the cavity as a part of the bath which is described by the Hamiltonian $H_\text{C} + H_\text{CB} + H_\text{B}$ including intra-bath interactions $H_\text{CB}$. The time-evolution of the full system-bath density matrix is governed by the von Neumann equation in the interaction picture
\begin{equation} \label{eq: von Neumann}
    \dot{\rho} \left( t \right) = - i \left[ H_\text{SB} \left( t \right), \rho \left( t \right) \right] =  \hat{\mathcal{L}}_\text{SB} \left( t \right) \rho \left( t \right)
\end{equation}
defining the Liouvillian superoperator $\hat{\mathcal{L}}_\text{SB} \left( t \right) = \left[ H_\text{SB} \left( t \right), \cdot \right]$ which is generically time-dependent in the interaction picture. We can insert the formal integral solution of Eq.~\ref{eq: von Neumann} back into Eq.~\ref{eq: von Neumann} to obtain
\begin{equation}
    \dot{\rho} \left( t \right) =  \hat{\mathcal{L}}_\text{SB} \left( t \right) \rho \left( 0 \right) + \int_0^t ds \; \hat{\mathcal{L}}_\text{SB} \left( t \right)  \hat{\mathcal{L}}_\text{SB} \left( s \right)  \rho \left( s \right) 
\end{equation}
which is second-order in the system-bath coupling strength.

From this starting point, we would like to obtain a tractable equation of motion for the reduced density matrix of the system $\rho_\text{S} \left( t \right) = \text{tr}_\text{B} \left( \rho \left( t \right) \right)$. To do so, we make standard assumptions on the factorization of the spin-bath density matrix and time-locality of the system-bath interactions \cite{breuer_theory_2002} which results in the Redfield master equation for the reduced density matrix of the system
\begin{equation}
    \dot{\rho}_\text{S} \left( t \right) = \int_0^t ds \; \text{tr}_\text{B} \left\{ \hat{\mathcal{L}}_\text{SB} \left( t \right)  \hat{\mathcal{L}}_\text{SB} \left( t - s \right)  \rho_\text{S} \left( t \right) \otimes \rho_\text{B} \right\}
\end{equation}
where $\rho_\text{B}$ is (usually) the equilibrium state of the bath. Finally, we assume that the time-scale of the systems dynamics is much longer than the decay of the bath correlation functions allowing us to extend the integration to the interval $\left[0, \infty \right]$ \cite{breuer_theory_2002}. In the remainder of the paper and the main text, we use the term ``Redfield master equation'' also to refer to this form of the quantum master equation, but before the secular approximation. 

In order to convert the Redfield master equation into a more useful more form, we decompose the system-bath interaction $H_\text{SB}$ as
\begin{equation}
    H_\text{SB} = \sum_k g_k S_k \otimes B_k
\end{equation}
in the Schrödinger picture, where $S_k$ are operators on the system, $B_k$ are operators on the bath, and $g_k$ are the couplings. Then the Redfield master equation becomes
\begin{equation} \label{eq: Redfield}
\begin{split}
    \dot{\rho}_\text{S} \left( t \right) & = - \sum_{k, k', \omega, \omega'} g_k g_{k'} e^{i \left( \omega + \omega' \right) t} \\
    & \quad \times \left\{ \tilde{C}^{(B)}_{kk'} \left( \omega', t \right) \left[ \tilde{S}_k \left( \omega \right), \tilde{S}_{k'} \left( \omega' \right) \rho_S \left( t \right) \right] + \tilde{C}^{(B)}_{k'k} \left( - \omega', - t \right) \left[ \tilde{S}_k \left( \omega \right), \rho_S \left( t \right) \tilde{S}_{k'} \left( \omega' \right) \right]\right\}
\end{split}
\end{equation}
where $\tilde{S}_k$ are the Fourier components of the system operators $S_k \left( t \right) = \sum_\omega e^{i \omega t} \tilde{S}_k \left( \omega \right)$ in the interaction picture, and the bath correlation functions 
\begin{equation}
    C^{(B)}_{kk'} \left( t \right) = \text{tr}_\text{B} \left( B_k \left( t \right) B_{k'} \left( 0 \right) \rho_\text{B} \right)
\end{equation}
are converted to frequency-space on a finite time-window of length $t$
\begin{equation} \label{eq: general correlation function in frequency}
    \tilde{C}^{(B)}_{kk'} \left( \omega', \pm t \right) = \int_0^{\pm t} ds \; e^{- i \omega' s} C^{(B)}_{kk'} \left( s \right)
\end{equation}
The sign of $t$ is crucial when extending the integration $t \to \infty$. 

Expanding the commutators in Eq.~\ref{eq: Redfield} and rearranging terms, we can obtain the master equation in the usual form in terms of a coherent Hamiltonian evolution and dissipative evolution. Additionally, the secular approximation is often used to neglect nonresonant transitions (terms with $\omega + \omega' \neq 0$ in Eq.~\ref{eq: Redfield}) induced by the system-bath coupling. However, we keep nonsecular terms in order to capture the superradiant behavior of the Dicke model, as discussed in the main text.

The adjoint Redfield master equation for a generic system operator $A$ is obtained from Eq.~\ref{eq: Redfield} by demanding that the expectation value of $A$ is unchanged between the Schrödinger and Heisenberg pictures \cite{breuer_theory_2002}. This results in the condition 
\begin{equation}
    \text{tr}_S \left( A \hat{\mathcal{L}}_\text{eff} \rho_\text{S} \left( t \right) \right) = \text{tr}_S \left( \left( \hat{\mathcal{L}}^*_\text{eff} A \left( t \right) \right) \rho_\text{S} \right)
\end{equation} 
where $\hat{\mathcal{L}}_\text{eff}$ is the effective Liouvillian superoperator governing the time-evolution of the reduced density matrix of the system $\dot{\rho}_\text{S} \left( t \right) = \hat{\mathcal{L}}_\text{eff} \rho_\text{S}  \left( t \right)$ in the Schrödinger picture, as in Eq.~\ref{eq: Redfield}, and $\hat{\mathcal{L}}_\text{eff}^*$ is the adjoint effective Liouvillian superoperator governing the time-evolution of correlation functions as $d \left\langle A \right\rangle / dt = \left\langle \hat{\mathcal{L}}_\text{eff}^* A \right\rangle$.

Having discussed general aspects of the derivation of the Redfield master equation, we apply the analysis to the system described in the main text. In particular, we note that generic system-bath interaction $H_\text{SB}$ used in the derivation is precisely the spin-cavity interaction $H_\text{SC}$ from the main text. Therefore we have the system operators $S_m = S_1^{(m)}$ and the bath operators $B_m = e^{- i \phi_m} a + e^{i \phi_m} a^\dagger$ with the couplings $g_m$. The Fourier decomposition of $S_1^{(m)}$ in the interaction picture is simply
\begin{equation}
    S_1^{(m)} \left( t \right) = \frac{1}{2} \left( e^{- i \epsilon t} S_-^{(m)} + e^{i \epsilon t} S_+^{(m)} \right)
\end{equation}
What remains to be calculated are the bath correlation functions, expanded as
\begin{equation}
\begin{split}
    C_{mm'}^{(B)} \left( t \right) & = e^{- i \left( \phi_m + \phi_{m'} \right)} \text{tr}_\text{B} \left( a \left( t \right) a \left( 0 \right) \rho_\text{B} \right) + e^{i \left( \phi_m + \phi_{m'} \right)} \text{tr}_\text{B} \left( a^\dagger \left( t \right) a^\dagger \left( 0 \right) \rho_\text{B} \right) \\
    & \quad + e^{- i \left( \phi_m - \phi_{m'} \right)} \text{tr}_\text{B} \left( a \left( t \right) a^\dagger \left( 0 \right) \rho_\text{B} \right) + e^{i \left( \phi_m - \phi_{m'} \right)} \text{tr}_\text{B} \left( a^\dagger \left( t \right) a\left( 0 \right) \rho_\text{B} \right)
\end{split}
\end{equation}
Assuming $\rho_\text{B}$ is an equilibrium thermal state, then we arrive at the standard expression
\begin{equation}
    C_{mm'}^{(B)} \left( t \right) = \int_{-\infty}^\infty \frac{ d \omega }{2 \pi} J_0^{(B)} \left( \omega \right) \left( e^{ i \left( \phi_m - \phi_{m'} \right)} \bar{n}_\text{eq} \left( \omega \right) e^{i \omega t} + e^{ - i \left( \phi_m- \phi_{m'} \right)} \left( \bar{n}_\text{eq} \left( \omega \right) + 1 \right) e^{- i \omega t} \right)
\end{equation}
modified due to the phases $\phi_m$, where $J_0^{(B)}$ is the zero-temperature spectral density formally given by the expression $J_0^{(B)} = 2 \pi \sum_j \left| c_j \right|^2 \delta \left( \omega - \omega_j \right)$ where the bath operator $B_m$ is expanded in the eigenmodes of $H_\text{C} + H_\text{CB} + H_\text{B}$ of energy $\omega_j$ with coefficients $c_k$, and $\bar{n}_\text{eq} \left( \omega \right)$ is the Bose-Einstein distribution function (at inverse temperature $\beta$) giving the thermal occupation of the bath eigenmode of energy $\omega$. In practice, specifying an approximate continuous zero-temperature spectral density enables analytic treatment of the correlation function, taking care to apply the Sokhotski–Plemelj theorem when applicable. In this paper, we assume the spectral density is Lorentzian (Cauchy) with a center frequency given by the cavity mode $\omega_c$ and half-width at half-maximum $\kappa$
\begin{equation}
    J_0^{(B)} = \frac{ 2 \kappa }{ \left( \omega - \omega_c \right)^2 + \kappa^2}
\end{equation}
which, like the exact form, is normalized to unity. Assuming zero or very low temperature, we obtained 
\begin{equation}
    C_{mm'}^{(B)} \left( t \right) = e^{- i \Delta \phi} e^{- i \omega_c t - i \kappa \left| t \right| }
\end{equation}
where $\Delta \phi = \phi_m - \phi_{m'}$ as presented in the main text, and 
\begin{align}
    \tilde{C}_{mm'}^{(B)} \left( \omega, t \right) & = e^{- i \Delta \phi} \left( \frac{ 1 - e^{- i \left( \omega_c + \omega \right) t - \kappa t}}{ i \left( \omega_c + \omega \right) + \kappa} \right) \\
    \lim_{t \to \infty} \tilde{C}_{mm'}^{(B)} \left( \omega, t \right) & = e^{- i \Delta \phi} \frac{1}{ i \left( \omega_c + \omega \right) + \kappa} \\
    \tilde{C}_{mm'}^{(B)} \left( \omega, - t \right) & = e^{- i \Delta \phi} \left( \frac{ 1 - e^{i \left( \omega_c + \omega \right) t - \kappa t}}{ i \left( \omega_c + \omega \right) - \kappa} \right) \\
    \lim_{t \to \infty} \tilde{C}_{mm'}^{(B)} \left( \omega, - t \right) & = e^{- i \Delta \phi} \frac{1}{ i \left( \omega_c + \omega \right) - \kappa}
\end{align}
Substitution of the above results into Eq.~\ref{eq: Redfield} yields Eq.~\ref{eq: spin only adjoint Redfield} or Eq.~\ref{eq: spin only adjoint Redfield alt}, discussed in the following section.

\subsection{Alternative form of the spin-only Redfield master equation}

As mentioned in the main text, we can transform Eq.~\ref{eq: spin only adjoint Redfield} into a form which highlights the connection to the mean-field equations of motion Eq.~\ref{eq: Redfield nonlinear mean-field EOM}. In particular, in terms of the Hermitian $x = 1$, $y = 2$, and $z = 3$ spin operators we have
\begin{equation} \label{eq: spin only adjoint Redfield alt}
    \begin{split}
    \hat{\mathcal{L}}^* A & = i \left[ H_\text{S}, A \right] + \sum_{m, m' = - M}^M \frac{ g_m g_{m'} }{ \sqrt{N_m N_{m'}} } \left( \left( \frac{ \cos \Delta \phi }{J_1} - \frac{ \sin \Delta \phi }{K_1} \right) \left[ \left[ S_1^{(m)}, A \right], S_1^{(m')} \right] \right. \\
    & \quad + \left( \frac{ \cos \Delta \phi }{J_2} - \frac{ \sin \Delta \phi }{K_2} \right) \left[ \left[ S_1^{(m)}, A \right], S_2^{(m')} \right] + \left( \frac{ \cos \Delta \phi }{K_1} + \frac{ \sin \Delta \phi }{J_1} \right) \left\{ - i \left[ S_1^{(m)}, A \right], S_1^{(m')} \right\} \\
    & \quad \left. + \left( \frac{ \cos \Delta \phi }{K_2} + \frac{ \sin \Delta \phi }{J_2} \right) \left\{ - i \left[ S_1^{(m)}, A \right], S_2^{(m')} \right\} \right) + \Gamma \sum_{m = - M}^M \sum_{j = 1}^{N_m} \hat{\mathcal{D}}^* \left[ s_-^{(m, j)} \right] A
    \end{split}
\end{equation}
with the parameters $J_1$, $J_2$, $K_1$, and $K_2$ defined in the main text. If $A$ is proportional to a product of $k$ spin operators, then the commutator-commutator terms in Eq.~\ref{eq: spin only adjoint Redfield alt} generically yield terms of the same order as $A$ which are sub-extensive and do not survive in the thermodynamic limit (though they may be significant for finite and small $N_m$). On the other hand, the anticommutator-commutator terms in Eq.~\ref{eq: spin only adjoint Redfield alt} generically yield symmetrized products of $k + 1$ spin operators (up to sub-extensive lower-order corrections) which are significant even in the thermodynamic limit. In the important case when $A$ is a first-order spin operator, the anticommutator-commutator terms in Eq.~\ref{eq: spin only adjoint Redfield alt} yield quadratic nonlinearities, as seen in Eq.~\ref{eq: Redfield nonlinear mean-field EOM}.

\section{Exact numerical diagonalization} \label{app: exact numerical diagonalization}

We study the system with low particle numbers by exact numerical diagonalization of the master equation after exploiting permutation symmetry. As noted in the main text, the density matrix can be decomposed into blocks of fixed spins $\left( s_i \right) = \left( s_{-M}, \ldots, s_M \right)$ as
\begin{align} 
    \rho & =  \sum_{\left( s_i \right)} p_{\left( s_i \right)} \rho_{\left( s_i \right)} \\
    & = \sum_{\left( s_i \right)} p_{\left( s_i \right)} \sum_{\left( \mu_i, \mu_i' \right)} p'_{\left( \left( \mu_i, \mu_i' \right) \right)} \left| \left( s_i, \mu_i \right) \right\rangle \left\langle \left( s_i, \mu_i' \right) \right| 
\end{align}
where $p_{\left( s_i \right)}$ is the probability of density matrix block $\rho^{\left( s_i \right)}$ which can be written explicitly in terms of the probabilities $p'_{\left( \left( \mu_i, \mu_i' \right) \right)}$ and the operator basis for fixed $\left( s_i \right)$, $\left| \left( s_i, \mu_i \right) \right\rangle \left\langle \left( s_i, \mu_i' \right) \right|$.
Here, $\mu_i$ is used for the magnetic quantum number of species $i$ in order to avoid confusion with the species label $m$. Then we vectorize the quantum master equation as $\left|\left. \dot{\rho} \right\rangle\right\rangle = \hat{\hat{\mathcal{L}}} \left|\left. \rho \right\rangle\right\rangle$, where $\left|\left. \rho \right\rangle\right\rangle$ is the vectorization of the density matrix $\rho$ and $\hat{\hat{\mathcal{L}}}$ is the vectorization of the Liouvillian superoperator in Lindblad form $\hat{\mathcal{L}}$, a matrix. Using the standard vectorization procedure, the vectorized density matrix \cite{fazio_many-body_2025} is
\begin{equation} 
    \left|\left. \rho \right\rangle\right\rangle = \sum_{\left( s_i \right)} p_{\left( s_i \right)} \sum_{\left( \mu_i, \mu_i' \right)} p'_{\left( \left( \mu_i, \mu_i' \right) \right)} \left|\left. \left( s_i, \mu_i, \mu_i' \right) \right\rangle\right\rangle
\end{equation}
For the vectorized Liouvillian superoperator in Lindblad form, we generalize the expressions in Ref.~\cite{shammah_open_2018} for off-diagonal dissipative processes (coherence between decay and pumping) and multiple species. The matrix elements are then
\begin{equation}
    \hat{\hat{\mathcal{L}}} = \sum_{\substack{ \left( s_i, \mu_i, \mu_i' \right) \\ \left( t_i, \nu_i, \nu_i' \right)}} \mathcal{L}_{\left( \left( s_i, \mu_i, \mu_i' \right), \left( t_i, \nu_i, \nu_i' \right) \right)} \left|\left. \left( s_i, \mu_i, \mu_i' \right) \right\rangle\right\rangle \left\langle\left\langle \left( t_i, \nu_i, \nu_i' \right) \right.\right|
\end{equation}
with explicit expressions for the matrix elements $\mathcal{L}_{\left( \left( s_i, \mu_i, \mu_i' \right), \left( t_i, \nu_i, \nu_i' \right) \right)}$ left out of this section for brevity. However, due to the nature of the Dicke coupling, excitations must be exchanged in pairs, thus constraining the matrix elements off-diagonal in $\mu_i$, $\mu_i'$, $\nu_i$, and $\nu_i'$. Additionally, the quantum jumps from single-particle incoherent decay (as well as single-particle dephasing and single-particle incoherent pumping) can only lead to coupling between total-spin manifolds satisfying $s_i = t_i$ or $s_i = t_i \pm 1$.

To visualize the density matrix, we first calculate the reduced density matrix of each species
\begin{equation}
    \rho^{(m)} = \text{tr}_{m' \neq m} \left( \rho \right)
\end{equation}
then write the reduced density matrix in block-diagonal form
\begin{equation}
     \rho^{(m)} = \sum_{s_m} p_{s_m}^{(m)} \rho^{(m, s_m)}
\end{equation}
Similar to the full density matrix, $p_{s_m}^{(m)}$ is the probability of density matrix block $\rho^{(m, s_m)}$.
Next, we extract the Wigner (quasi-probability) distribution for species $m$ of fixed spin $s_m$ as
\begin{equation}
    W_{s_m}^{(m)} \left( \vartheta, \varphi \right) = \text{tr} \left( \mathcal{W}_{s_m} \left( \vartheta, \varphi \right) \rho^{(m, s_m)} \right)
\end{equation}
where
\begin{equation}
    \mathcal{W}_s \left( \vartheta, \varphi \right) = \sqrt{\frac{4\pi}{2 s + 1}} \sum_{\lambda = 0}^{2 s} \sum_{\mu = - \lambda}^\lambda Y_{\lambda \mu} \left( \vartheta, \varphi \right) T_{\lambda \mu}^{(s) \dagger}
\end{equation}
is the Wigner (operator) kernel, $Y_{\lambda \mu}$ are spherical harmonics, and $T_{\lambda \mu}^{(m, s)}$ are irreducible tensor operators \cite{klimov_group-theoretical_2009}. We note that the Wigner kernel has the following properties
\begin{gather}
    \mathcal{W}_s\left( \vartheta, \varphi \right) = \mathcal{W}_s \left( \vartheta, \varphi \right)^\dagger \\
    \text{tr} \left( \mathcal{W}_s\left( \vartheta, \varphi \right) \right) = 1 \\
    \frac{2 s + 1}{4 \pi} \int d \Omega \: \mathcal{W}_s \left( \vartheta, \varphi \right) = 1
\end{gather}
Since the Wigner kernel is Hermitian (as is the density matrix), the Wigner distribution is real. Moreover, it can be normalized, though permitting negative values. Lastly, we average over the allowed values of spin $s_m$ to obtain the spin-averaged Wigner distribution. 
\begin{equation}
    W^{(m)} \left( \vartheta, \varphi \right) = \sum_{s_m} p_{s_m} \left( \frac{2 s_m + 1}{4 \pi} \right) W_{s_m}^{(m)}\left( \vartheta, \varphi \right)
\end{equation}
If the reduced density matrix $\rho^{(m)}$ is proportional to the identity, i.e. a completely mixed state, then the Wigner distribution for species $m$ of fixed spin $s_m$ is simply
\begin{equation}
    W_{s_m}^{(m)} \left( \vartheta, \varphi \right) = \frac{1}{2 s_m + 1}
\end{equation}
and the spin-averaged Wigner distribution is $W^{(m)} \left( \vartheta, \varphi \right) = 1 / 4 \pi$. This value $1 / 4 \pi \approx 0.08$ is used in the main text as a reference.

\end{appendix}

\bibliography{bibliography}

\end{document}